%% file: main.tex
\documentclass[sigconf]{acmart}

\AtBeginDocument{%
  }

\copyrightyear{2024}
\acmYear{2024}
\setcopyright{acmlicensed}\acmConference[CCS '24]{Proceedings of the 2024
ACM SIGSAC Conference on Computer and Communications Security}{October
14--18, 2024}{Salt Lake City, UT, USA}
\acmBooktitle{Proceedings of the 2024 ACM SIGSAC Conference on Computer
and Communications Security (CCS '24), October 14--18, 2024, Salt Lake City,
UT, USA}
\acmDOI{10.1145/3658644.3690367}
\acmISBN{979-8-4007-0636-3/24/10}

\usepackage{multirow}
\usepackage{enumitem}
\usepackage[linesnumbered,ruled,noline,nofillcomment]{algorithm2e}
\usepackage{empheq}
\usepackage{soul}
\usepackage{bbm}
\usepackage{color}
\usepackage{graphicx}
\usepackage{subcaption}

\newtheorem{definition}{Definition}
\newtheorem{remark}{Prerequisite}
\newtheorem{theorem}{Theorem}

\begin{document}

\title{Curator Attack: When Blackbox Differential Privacy Auditing Loses Its Power}

\author{Shiming Wang}
\affiliation{%
  \institution{Shanghai Jiao Tong University}
  \city{Shanghai}
  \country{China}
}
\email{my16wsm@sjtu.edu.cn}

\author{Liyao Xiang$^{\dag}$}\thanks{$^{\dag}$Corresponding author}
\affiliation{%
  \institution{Shanghai Jiao Tong University}
  \city{Shanghai}
  \country{China}
}
\email{xiangliyao08@sjtu.edu.cn}

\author{Bowei Cheng}
\affiliation{%
  \institution{Shanghai Jiao Tong University}
  \city{Shanghai}
  \country{China}
}
\email{chengbowei@sjtu.edu.cn}

\author{Zhe Ji}
\affiliation{%
  \institution{Shanghai Jiao Tong University}
  \city{Shanghai}
  \country{China}
}
\email{ji_zhe@sjtu.edu.cn}

\author{Tianran Sun}
\affiliation{%
  \institution{Shanghai Jiao Tong University}
  \city{Shanghai}
  \country{China}
}
\email{Seriousss@sjtu.edu.cn}

\author{Xinbing Wang}
\affiliation{%
  \institution{Shanghai Jiao Tong University}
  \city{Shanghai}
  \country{China}
}
\email{xwang8@sjtu.edu.cn}

\begin{abstract}
  A surge in data-driven applications enhances everyday life but also raises serious concerns about private information leakage. Hence many privacy auditing tools are emerging for checking if the data sanitization performed meets the privacy standard of the data owner. Blackbox auditing for differential privacy is particularly gaining popularity for its effectiveness and applicability to a wide range of scenarios. Yet, we identified that blackbox auditing is essentially flawed with its setting --- small probabilities/densities are ignored due to inaccurate observation. Our argument is based on a solid false positive analysis from a hypothesis testing perspective, which is missed out by prior blackbox auditing tools. This oversight greatly reduces the reliability of these tools, as it allows malicious or incapable data curators to pass the auditing with an overstated privacy guarantee, posing significant risks to data owners. We demonstrate the practical existence of such threats in classical differential privacy mechanisms against four representative blackbox auditors with experimental validations. Our findings aim to reveal the limitations of blackbox auditing tools, empower the data owner with the awareness of risks in using these tools, and encourage the development of more reliable differential privacy auditing methods.
\end{abstract}




\maketitle
\renewcommand{\shortauthors}{Shiming Wang et al.}
\input{1_intro.tex}

\input{3_problem_and_threat.tex}

\input{4_motivation.tex}

\input{5_roadmap.tex}

\input{6_sniper}

\input{7_mpl.tex}
\input{8_delta-siege.tex}
\input{9_dpsgd.tex}
\section{Discussion}

\textbf{Constructing a curator attack.} We show how to construct a successful curator attack with our previous false positive analysis.
As an example, consider a malicious curator who intends to use the SVT mechanism, but pass DP-Sniper's auditing with an overstated $\epsilon_c$-DP claim. To launch this attack, the curator could first solve Eq. (\ref{eq:svt_sniper_benchmark}), select any $\theta$ from the solution, and deploy the benchmark SVT mechanism with this $\theta$. If Eq. (\ref{eq:svt_sniper_benchmark}) has no solution, the curator could tweak the mechanism as in Alg. \ref{alg:altered_svt}, solve Eq. (\ref{eq:svt_sniper}) instead, and use any  $\tilde{\theta}$ from its solution to deploy the adapted SVT mechanism.
The curator could now claim to satisfy $\epsilon_c$-DP without getting caught. Attacks involving other blackbox auditors and mechanisms can be constructed in the same way.

\textbf{Confidence interval of $\xi^*$.}
When computing $\xi^*$ in our analysis above, we directly use the theoretical $\Pr[M(a)\in S]$ and $\Pr[M(a')\in S]$ for large values.
The implicit assumption is that the auditor's estimation of large probabilities above $c$ is accurate. Hence our analytical computation of $\xi^*$ represents the theoretical value of the auditor's empirical power.

In practice, these probabilities are also estimated via randomized sampling, which introduces slight variations across multiple auditing runs due to sampling randomness. 
Therefore, existing auditors typically report a confidence interval for its power instead of a single value.
Specifically, DP-Sniper \cite{snipercode}, MPL \cite{askin2022statistical} and Delta-Siege \cite{deltasiege} employs a one-sided confidence interval $[\underline{\xi^*},\infty)$. DPSGD-Audit \cite{nasr2023tight} reports a two-sided interval $[\underline{\xi^*},\overline{\xi^*})$ \cite{zanella2023bayesian}.

A privacy claim $\epsilon_c$ passes the auditing if and only if it falls within the auditor's confidence interval. Hence for our evaluation in \S\ref{sec:experiments}, we only need to report the lower bound $\underline{\xi^*}$ for all auditing tools except for DPSGD-Audit.

\textbf{Dilemma between auditor's reliability \& applicability.}
A natural question following our discussion is the future attempts towards better blackbox auditing. Current tools are unreliable because they fail to examine the small probability region.
Hence to eliminate FPs universally, blackbox auditing must inspect the entire type II-type I tradeoff curve of the curator.

One potential move is to ``open the blackbox" and leverage the mechanism's inner structure to parametrically estimate the tradeoff, as suggested in \cite{nasr2023tight}. For instance, if the curator uses Gaussian noise, then its tradeoff curve has the parametric form of $\mu$-Gaussian differential privacy \cite{dong2019gaussian}. With this knowledge, the auditor could estimate the parameter $\mu$ from the collected samples and obtain the entire tradeoff fairly accurately, including the small probability segments, thereby resolving the false positive issue.

However, this approach violates the blackbox auditing assumption, which limits the tool's applicability fundamentally. It cannot audit any curator mechanisms that are proprietary or have intricate tradeoff functions (e.g. the SVT mechanism).
Therefore, blackbox auditors are stuck in this dilemma between reliability and applicability for the moment, and perfecting them remains an open problem.

\textbf{Extending to other auditors.}
Our analysis of FP and FN can be generalized to future blackbox auditors. We first extend the prerequisites of FP and FN as follows, and the rest of the analysis pipeline is the same as Alg \ref{alg:roadmap}.

Step 1: Given any blackbox auditor, we traverse the outcome sets of the curator mechanism and mathematically formulate the auditor's error in estimating the curator's theoretical probability on each outcome set. For current blackbox auditors, this corresponds to concluding the pattern of ``omitting small probabilities'' in \S\ref{sec:sample}.

Step 2: With this formulation, given any benchmark DP mechanism, we could explicitly deduce the auditor's observed tradeoff $\mathcal{T}_{audit}$ and identify the ``region of discrepancy", where the observed $\mathcal{T}_{audit}$ differs from the curator's theoretical tradeoff $\mathcal{T}_{curator}$. 
For example, for existing blackbox auditors, this is the small probability region marked in Figure \ref{fig:tradeoff} (c)(d).

Step 3: Given a claimed privacy parameter $\delta_c$, we identify the theoretical optimal witness from the theoretical tradeoff (eg. Eq. \eqref{eq:optimal_outcome_ratio}).
Then we tweak the curator mechanism so that the optimal witness falls within the region of discrepancy.
For example, in \S\ref{sec:lap_sniper}, we adjust the Laplace mechanism's $\theta$ or adapt its noise distribution to $\widetilde{\text{Lap}}(\Delta/\tilde{\theta})$, so that the optimal $S^*$ lies within the small probability region as in Eq. \eqref{eq:sniper_if}. The resulting mechanism will be a false positive or false negative in our analysis.

\section{Experiments}
\label{sec:experiments}
In this section, we empirically verify that all identified benchmark and adapted mechanisms in previous sections are indeed false positives (negatives), suggesting possible curator attacks. 
\subsection{Setup}
\textbf{DP curator mechanisms.}
We first act as the data curator and implement all benchmark and adapted mechanisms in \texttt{Python}. The DP mechanisms under examination include Laplace, SVT \cite{svt}, one-time RAPPOR \cite{rappor}, Gaussian, and DPSGD. We move the results on one-time RAPPOR to Appendix J \cite{appendix}. 

For each benchmark mechanism $M_{\theta}$, we traverse the possible values of its parameter $\theta$, derive $\epsilon^*(\theta)$ by definition, and perform corresponding auditing to obtain an empirical $\xi^*(\theta)$.
For an intuitive and comprehensive revelation of FPs, each benchmark $M_{\theta}$ is illustrated with a plot of the relationship between $\epsilon^*(\theta)$ and $\xi^*(\theta)$. For each adapted mechanism $M_{\widetilde{\theta}}$, we show how the adapted $\widetilde{\theta}$ leads to FPs provided a range of privacy claim $\epsilon_c$.

\textbf{Auditing tools.}
For a fair evaluation, we reuse the official code in \cite{bichsel2021dp}, \cite{askin2022statistical} and \cite{deltasiege} to eliminate any potential faulty implementations. 
For DP-Sniper, MPL, and Delta-Siege, the adjacent datasets are constructed using the \texttt{GenerateInputs()} function in \cite{bichsel2021dp}, which follows the input pattern heuristic as summarized in Table \ref{tab:heuristic}.
Specifically, the sampling numbers of DP-Sniper and MPL are set according to the instructions in the original work of DP-Sniper and MPL (\cite{bichsel2021dp} Theorem 2 and \cite{askin2022statistical} (C3) and (D)).
Specifically, the sampling number of DP-Sniper is $10.7\cdot 10^{6}$ for $c=0.01$ and $2.05\cdot 10^6$ for $c=0.05$. The sampling number of MPL is $3\cdot 10^6$.
 We report the one-sided 0.95 confidence interval of $\xi^*$.  The sampling number of Delta-Siege is 15000, and we report the one-sided  0.9 confidence interval of $\xi^*$.

For DPSGD-Audit, we evaluate the tasks of image classification and next word prediction on datasets and models as in Table \ref{tab:example}. The sampling number is $N=1000$ for CIFAR10, and $N=10000$ for SVHN and WikiText.
We report the two-sided bayesian interval \cite{zanella2023bayesian} of $\xi^*$  with siginificance level below 0.03, which is the superior confidence interval for DPSGD auditing so far.

\small
\begin{table}[h]
	\centering
	\caption{Tasks, datasets, models and parameters of the DPSGD-Audit.}
	\resizebox{0.98\linewidth}{!}{
	\begin{tabular}{p{0.279\linewidth}|p{0.19\linewidth}|p{0.13\linewidth}|p{0.14\linewidth}|p{0.05\linewidth}}
	\toprule
	DPSGD Task & Dataset & Model & Sampling number $N$ & $\delta_c$ \\
	\midrule
	Image
	& CIFAR10 & \multirow{2}{*}{ResNet20} & 1000 & \multirow{2}{*}{$10^{-4}$} \\
	classification					 & SVHN &  & 10000 &  \\
	\midrule
	Next word prediction & WikiText-103 & T5 & 10000 & $10^{-5}$ \\
	\bottomrule
	\end{tabular}
	}
	\label{tab:example}
	\end{table}
	\normalsize

\small
\begin{table}[t]
	\centering
	\caption{Input patterns adopted from \cite{bichsel2021dp} and \cite{askin2022statistical} (using 5-dimensional output as an example).}
	\label{tab:heuristic}
	\begin{tabular}{l|c|c}
		\toprule
		Pattern & Query $q(a)$ &Query $q(a')$\\
		\midrule
		One Above & [1,1,1,1,1] & [2,1,1,1,1] \\
		One Below & [1,1,1,1,1] & [0,1,1,1,1] \\
		One Below Rest Above  & [1,1,1,1,1] & [0,2,2,2,2] \\
		Half Half & [1,1,1,1,1] & [0,0,0,2,2] \\
		All Above \& All Below & [1,1,1,1,1] & [2,2,2,2,2] \\
		X shape & [1,1,0,0,0] & [0,0,1,1,1]\\
		\bottomrule
	\end{tabular}
\end{table}
\normalsize

\subsection{False Positives against DP-Sniper}
\begin{figure}[t]
	\centering
	\includegraphics[width=0.95\linewidth]{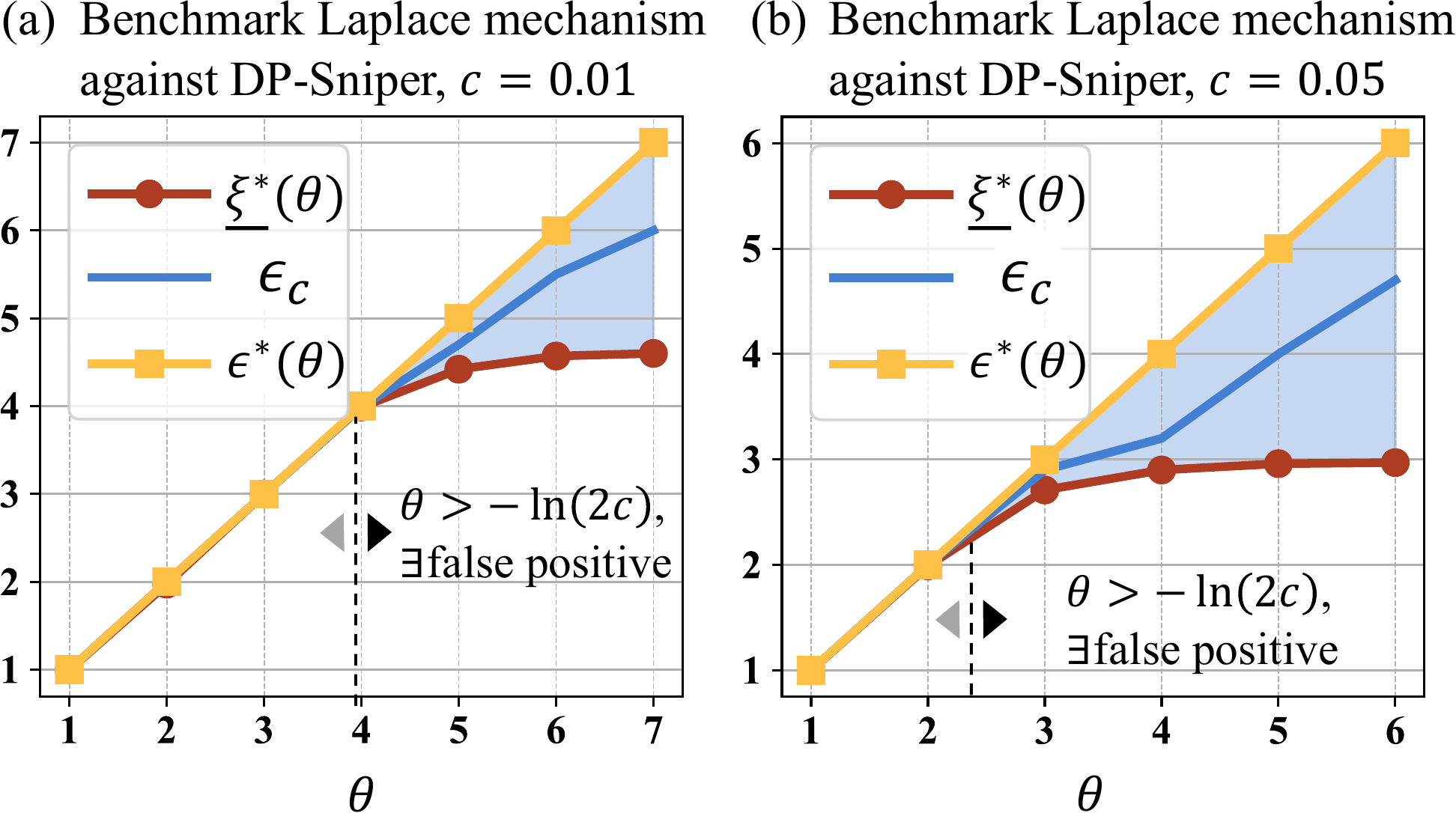}
	\caption{The benchmark Laplace mechanism is an FP against DP-Sniper at $\theta>-\ln(2c)$.}
	\label{fig:result_lap_benchmark}
\end{figure}

\textbf{Laplace mechanism.} We set the probability threshold as $c=0.01$ or $c=0.05$. As illustrated in Figure \ref{fig:result_lap_benchmark}, when $\theta>-\ln(2c)$, there is a gap between the mechanism's privacy level and DP-Sniper's empirical power. Therefore, any privacy claim $\epsilon_c$ within the blue region makes the benchmark Laplace mechanism a false positive, aligning with Thm. \ref{theorem:laplace_tight}. Meanwhile, for $\theta < -\ln(2c)$, the auditing is validated to be tight, reflecting the iff condition.

\begin{figure}[t]
	\includegraphics[width=\linewidth]{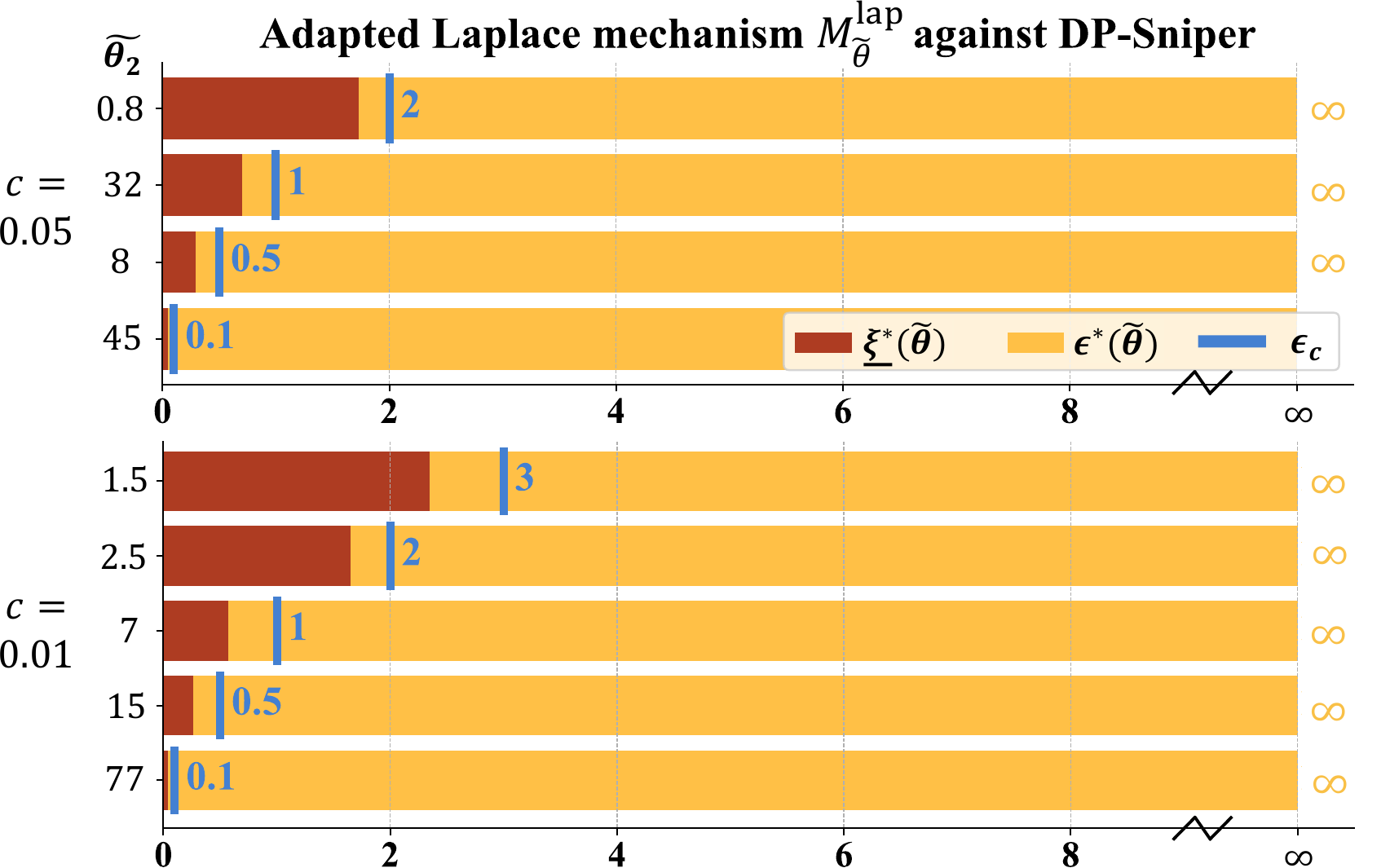}
	\caption{Adapted Laplace mechanism against DP-Sniper's auditing, $c=0.01$ or $0.05$. All adapted mechanisms are FPs.}
	\label{fig:result_sniper_lap_adapted}
\end{figure}

For the adapted Laplace mechanism, we set $\epsilon_c\in \{0.1, 0.5, 1, 2\}$ for $c=0.01$ and $c=0.05$.
We first solve $\widetilde{\theta}$ following Thm. \ref{theorem:adapted_lap_sniper}. To satisfy $\widetilde{\theta_1}<\epsilon_c$ in Eq. \eqref{eq:adapted_lap_sniper}, w.l.o.g., we let $\widetilde{\theta_1}=\epsilon_c/2$. We then arbitrarily choose an $\widetilde{\theta}_2$ from the solution set as shown in Figure \ref{fig:result_sniper_lap_adapted}. The theoretical privacy level is $\epsilon^*(\widetilde{\theta})=\infty$ as marked yellow, indicating the adapted Laplace mechanisms do not satisfy any DP. Yet they all pass the auditing as DP-Sniper's power (marked red) is lower than the $\epsilon_c$-DP privacy claim (marked blue), making them false positives.

\begin{figure}
	\centering
	\includegraphics[width=0.95\linewidth]{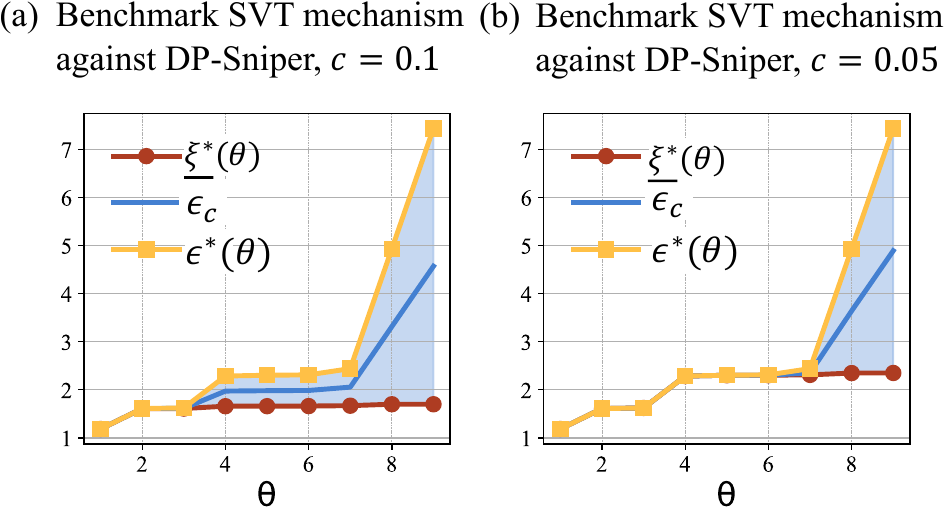}\caption{The benchmark SVT against DP-Sniper's auditing.}
	\label{fig:result_sniper_svt_benchmark}
\end{figure}

\textbf{SVT.} For the benchmark SVT mechanism, we set $T=1$, $N=1$ and $\bar{t}=1$. As illustrated in Figure \ref{fig:result_sniper_svt_benchmark}, the benchmark SVT behaves similarly to the Laplace mechanism. The curator can choose any privacy claim within the blue region without being detected. 

\begin{figure}[t]
	\includegraphics[width=\linewidth]{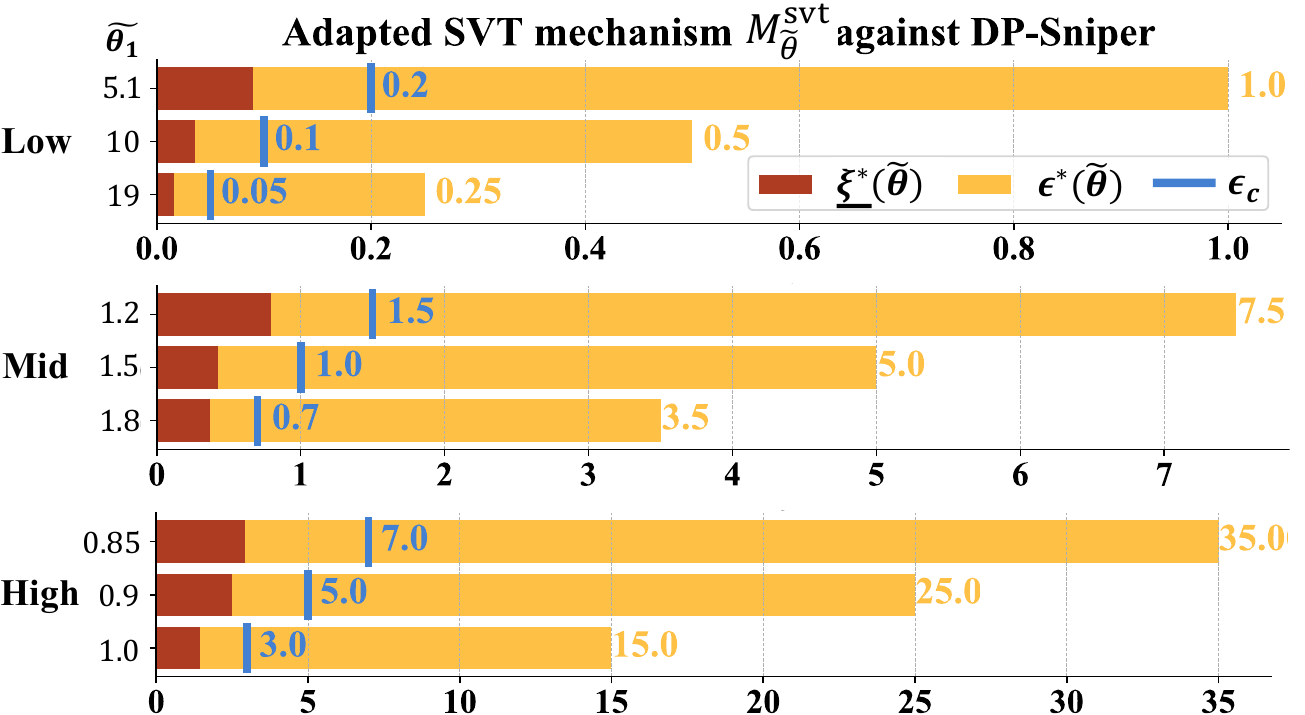}
	\caption{Adapted SVT mechanism against DP-Sniper's auditing, $c=0.01$. }
	\label{fig:result_adaptedsvt_sniper}
\end{figure}

For our adapted SVT mechanism, we set $T=1$, $N=10$ and $\bar{t}=1$ . We select a wide range of $\epsilon_c$ values, encompassing weak, moderate, and strong privacy claims shown in Figure \ref{fig:result_adaptedsvt_sniper}. Notice that the yellow bar is not the precise $\epsilon^*(\widetilde{\theta})$ but one of its lower bounds, as introduced in \S\ref{sec:svt_sniper}. Therefore, the actual gap between DP-Sniper's auditing and the adapted mechanism's actual privacy level can be even larger than illustrated. Hence all adapted mechanisms are false positives.

\subsection{False Positives against MPL} 
\begin{figure}
	\includegraphics[width=\linewidth]{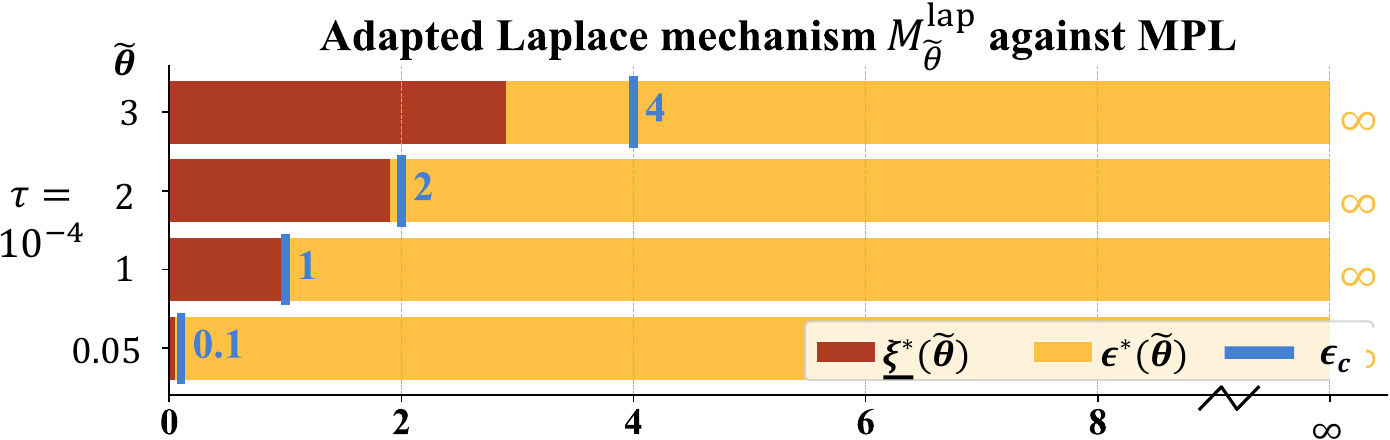}
	\caption{Adapted Laplace mechanism against MPL's auditing, $\tau=10^{-4}$.}
	\label{fig:result_mpl_lap}
\end{figure}

\begin{figure}
	\includegraphics[width=\linewidth]{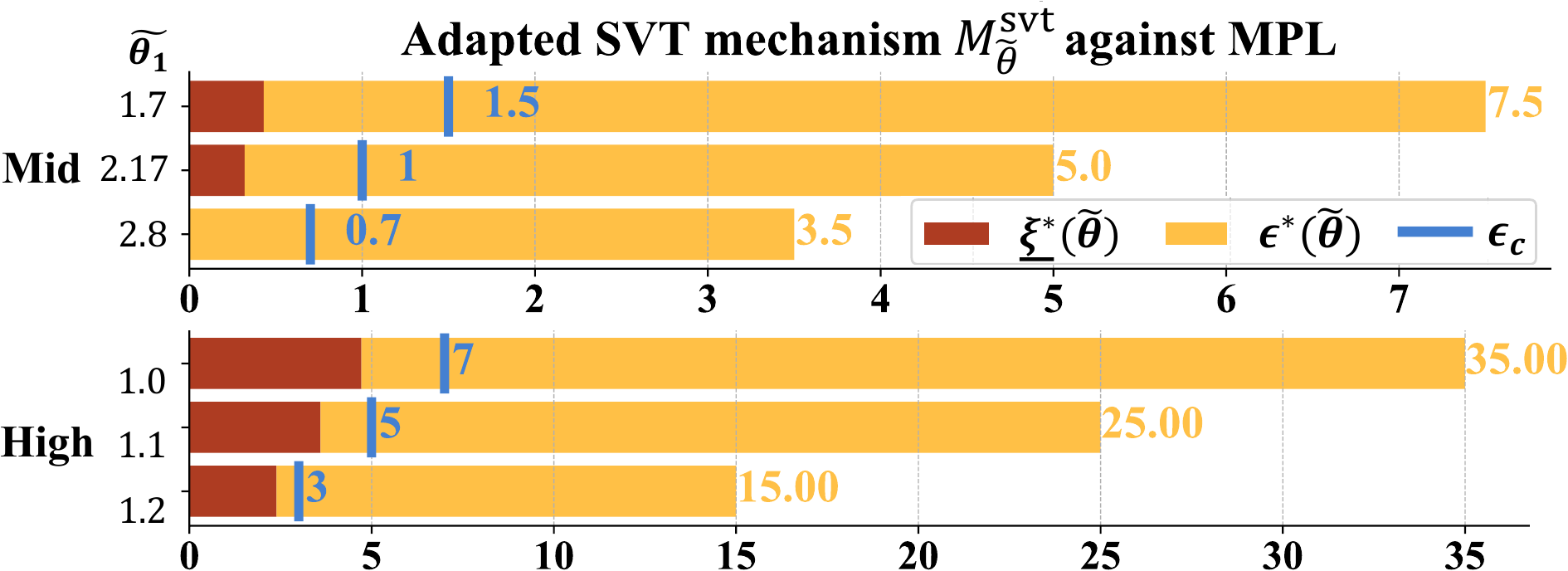}
	\caption{Adapted SVT against MPL's auditing, $\tau=10^{-4}$. Results in the low parameter regions are omitted because MPL's power $\xi^*$ are almost 0.}
	\label{fig:result_mpl_svt}
\end{figure}

As discussed in \S\ref{sec:fp_mpl}, we only need to discuss FP's occurrence in MPL against the adapted Laplace mechanism, the adapted SVT mechanism, and the benchmark RAPPOR mechanism. The conclusions are similar to that of DP-Sniper, and the results are in Figure \ref{fig:result_mpl_lap}, \ref{fig:result_mpl_svt} and Appendix J \cite{appendix}.


\subsection{FPs \& FNs against Delta-Siege}
The auditing results of Delta-Siege are empirically unstable between multiple audit runs. Hence we follow its original work and run five independent audit runs for each $\theta$. The reported $\xi^*(\theta)$ is the maximum among all runs. As shown in Table \ref{tab:siege_result}, for the same privacy surrogate $\rho(\epsilon,\delta)$, the Gaussian mechanism can be a false positive for some $\theta$ and a false negative for others. Hence Delta-Siege's auditing is unreliable under a blackbox setting.
Results of the Laplace mechanism are omitted as they are similar to that of DP-Sniper. 

\small
\begin{table}[ht]
	\renewcommand{\arraystretch}{1.25}
	\centering
	\caption{Gaussian mechanism $M^{N}_{\theta}$ against Delta-Siege auditing, with a privacy surrogate $\rho(\epsilon,\delta)=1/(e^{\epsilon}\delta)$.}
	\label{tab:siege_result}
	\begin{tabular}{l|c|cc|c}
		\toprule
		$\rho(\epsilon,\delta)=\dfrac{1}{e^{\epsilon}\delta}$ & Empirical optimal $({\alpha},{\beta})$ &$\epsilon^*(\theta)$ & $\xi^*(\theta)$ & Error
		\\
		\midrule
		\multirow{2}{*}{$\delta_c=0.005$} & $(0.055,0.921)$ & $0.30$ & $1.56$ & FN \\
		& $(0.0006,0.975)$ & $3.5$ & $3.9$ & FN \\
		\midrule
		$\delta_c=0.05$ & $(0.005,0.675)$ & $5.1$ & $4.7$ & FP\\
		\bottomrule
	\end{tabular}
\end{table}
\normalsize

\subsection{False Positives against DP-SGD Audit}
\begin{figure}
	\includegraphics[width=0.85\linewidth]{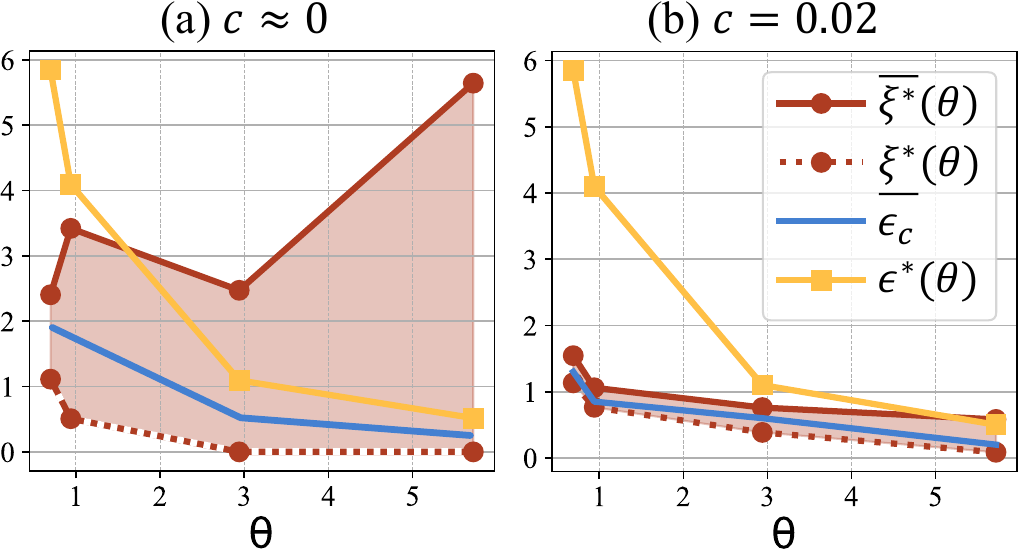}
	\caption{Blackbox DPSGD auditing,  two-sided bayesian confidence interval $\xi^*$.  $\delta=10^{-4}$. Image classification task on CIFAR-10 with ResNet-20.}
	\label{fig:result_dp_sgd}
\end{figure}

\begin{figure}
	\centering
	\includegraphics[width=0.85\linewidth]{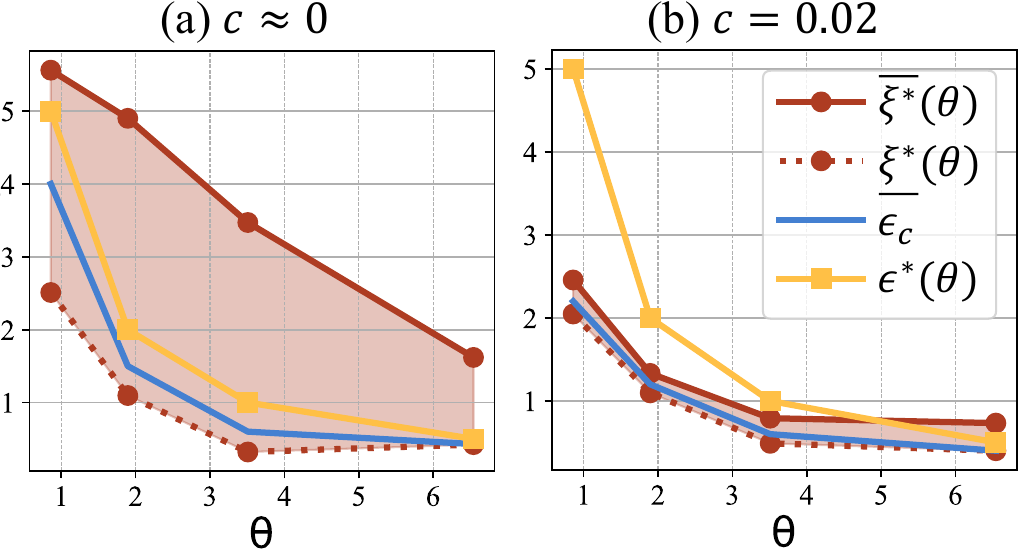}
	\caption{Blackbox DPSGD auditing,  two-sided bayesian confidence interval $\xi^*$.  $\delta_c=10^{-4}$. Image classification task on SVHN with ResNet-20.}
	\label{fig:result_dp_sgd_svhn}
\end{figure}

\begin{figure}
	\includegraphics[width=0.85\linewidth]{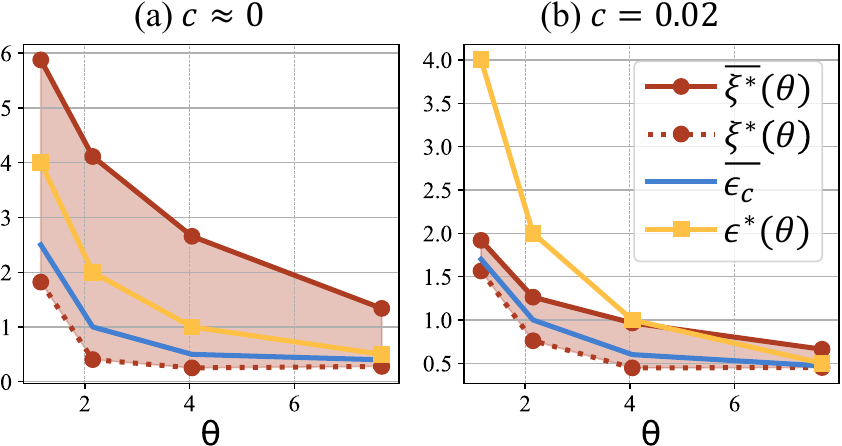}
	\caption{Blackbox DPSGD auditing,  two-sided bayesian confidence interval $\xi^*$.  $\delta=10^{-5}$. Next word prediction task on WikiText-103 with T5.}
	\label{fig:result_dp_sgd_wiki}
\end{figure}

We report the two-sided confidence interval $(\underline{\xi^*},\overline{\xi^*})$ of the DPSGD auditor's power with a significance level below 0.03 \cite{zanella2023bayesian}. As illustrated in Figure \ref{fig:result_dp_sgd}, \ref{fig:result_dp_sgd_svhn} and \ref{fig:result_dp_sgd_wiki}, any $\epsilon_c$ claim that falls within the red confidence interval and below the yellow $\xi^*(\theta)$ serves as a false positive.

Selecting extreme threshold values allows the auditor to obtain very small probabilities, i.e. $c\approx 0$. However, as discussed in \S\ref{sec:sample}, such probability estimations are highly unstable, which leads to a confidence interval too wide to be useful because it allows numerous false positive $\epsilon_c$ claims (Figure \ref{fig:result_dp_sgd}(a), \ref{fig:result_dp_sgd_svhn}(a) and \ref{fig:result_dp_sgd_wiki}(a)). For a comprehensive analysis, we also report the results when the auditor uses moderate thresholds instead, resulting in $c=0.02$. As shown in Figure \ref{fig:result_dp_sgd}(b), \ref{fig:result_dp_sgd_svhn}(b) and \ref{fig:result_dp_sgd_wiki}(b), the confidence interval is now significantly narrower and hence more informative. However, the $\epsilon_c$ claims are also false positives as the confidence interval is still much lower than $\epsilon^*(\theta)$. 




\input{2_related.tex}
\section{Conclusion}While blackbox auditing for differential privacy has become popular for its wide applicability, our research reveals a critical flaw: the neglect of small probabilities/densities due to imprecise observation. This gap, highlighted through our false positive analysis, compromises the reliability of these tools, allowing data curators with exaggerated privacy claims to go undetected. Our practical experiments with classical differential privacy mechanisms further underscore the risks posed by such curators. Ultimately, our work sheds light on the vulnerabilities in current blackbox auditing practices, aims to heighten data owners' awareness of these risks, and calls for the advancement of more dependable differential privacy auditing solutions.

\begin{acks}
  The research was supported in part by National Science and Technology Major Project 2021ZD0112801, NSF China (62272306, 62032020, 62136006).
\end{acks}

\bibliographystyle{ACM-Reference-Format}
\bibliography{main}

\end{document}

%% file: 1_intro.tex
\section{Introduction}

The rise of data-driven applications, e.g., healthcare data and software usage analysis, hinges upon the massive scale of data collected from abundant sources. Yet this data usually lacks close inspection and is at risk of severe (personal) information leakage. A notable instance is the potential leakage of sensitive personal data in large language model-based services, as reported by Kim et al. \cite{kim2024propile}. Since ordinary \textit{data owners} lack powerful means to safeguard their data independently, they entrust their data to a \textit{data curator}.
These curators are responsible for sanitizing the data prior to its release or removing private information in a preprocessing step, as outlined in several studies  \cite{rappor,ding2017collecting,johnson2018towards,abowd2018us}. In particular, if the data is protected by differential privacy (DP) --- the golden standard for restricting record-level privacy leakage --- the curator is expected to apply DP mechanisms by the privacy level designated by the data owner.

Hereby data owners often raise concerns about whether the sanitized data truly meets the required privacy level, as the curator can be malicious or incapable, failing to deliver the promised privacy and putting the data owners at risk. Hence, a third-party \emph{DP auditor} plays an important role in ascertaining whether the curator properly implants DP mechanisms at a sufficient level of privacy. This motivates a line of auditing solutions including  \cite{bichsel2018dp,ding2018detecting,bichsel2021dp,askin2022statistical,ravi2019automated,wang2020checkdp,dixit2013testing,wilson2020differentially,barthe2013verified,barthe2014proving,barthe2016advanced,liu2019minimax,barthe2020deciding} for $\epsilon$-DP, as well as $(\epsilon,\delta)$-DP  \cite{barthe2013verified,barthe2014proving,barthe2016advanced,liu2019minimax,barthe2020deciding,deltasiege}.

Among various auditing solutions, \emph{blackbox auditing} is becoming the dominant approach for its effectiveness and extensive applicability. Blackbox auditing refers to a scheme that seeks the most powerful witness with mere blackbox access to the DP mechanism. However, lacking an internal view of the audited DP mechanism, blackbox auditing can hardly find the optimal witness in practice. It often relies on the output samples of the queries to find a surrogate witness. Such an auditing imposes the least requirement on the data curator as no raw data or the internal workings of the mechanism are revealed, which is particularly desirable when the mechanism runs on proprietary or complex algorithms.

Representative blackbox auditing tools \cite{bichsel2021dp,askin2022statistical,deltasiege} typically make a non-parametric estimation of the privacy level given collected output samples. The sampling-based approach inevitably leads to significant error in the estimation of small probabilities or densities. In response, DP-Sniper \cite{bichsel2021dp} and MPL \cite{askin2022statistical} proactively constrain the estimation within the large probability (density) regions and ignore small probabilities (densities). Delta-Siege \cite{deltasiege} samples a surrogate variable that largely circumvents the small probability events. However, we observe the error in small probability estimation persists and eventually causes blackbox auditing schemes to fail to catch the overclaimed privacy of malicious curators.



In this paper, we re-examine the pitfall of existing blackbox auditing tools. Our analysis does not entirely negate blackbox auditing; rather it systematically identifies the inherent weakness of blackbox auditing, and empowers the data owners with the awareness of the risk of using existing auditing services. Our argument is based on a solid evaluation of these tools through the lens of hypothesis testing: the null hypothesis $H_0$ is ``the data curator's $\epsilon_c$-DP claim is valid,'' whereas the alternative hypothesis $H_1$ is ``the curator's provided privacy is weaker than claimed.''  
This novel formulation describes our contribution as \textit{auditing the effectiveness of auditing tools}, which should not be confused with the canonical hypothesis testing interpretation of differential privacy \cite{dong2019gaussian} that \textit{audits the effectiveness of DP mechanisms}, as elaborated in \S\ref{sec:motivation}.

High FPs indicate the auditor fails to catch curators with exaggerated privacy claims, posing privacy leakage threats to data owners. 
By our hypothesis testing, we find that false positives (FPs) widely exist in most blackbox auditing tools against both benchmark and adapted curator mechanisms. We conclude that \emph{false positives are ubiquitous in blackbox auditing because it ignores small probabilities (densities)}. As blackbox auditors limit their computation within large probabilities (densities), they can inspect only a segment of the curator's type II error - type I error tradeoff curve on the canonical hypothesis testing interpretation of DP. In the remaining portion beyond the auditors' scrutiny, curators can manipulate their curves at will. This unchecked flexibility permits undetected privacy violations, leading to a surge in FPs.

Beyond FPs, false negatives (FNs) are also spotted in certain blackbox auditors, which suggest the auditor would wrongly refute a valid privacy claim. This leads to mistrust in curators who offer adequate privacy protection. The frequent occurrence of FPs and FNs should raise significant concerns in blackbox auditing, even if the auditors reliably validate some correct DP claims (high true positives). Our findings are summarized in Table~\ref{tab:organization}.

False positives (FPs) in auditing are particularly fatal, as they compromise the privacy of data owners directly. To demonstrate the privacy impact of FPs in real-world applications, we translate them into realistic \emph{curator attacks}. In these attacks, curators aim to secretly divulge the owners' private data, by providing deceptive DP mechanisms to curate the data while evading the auditors' detection. Hereby we introduce three executable conditions to create such attacks, instantiate these conditions by specific benchmark/adapted DP mechanisms, and show how they evade the state-of-the-art blackbox auditing tools including DP-sniper \cite{bichsel2021dp}, MPL \cite{askin2022statistical}, and Delta-siege \cite{deltasiege}. We observe that this FP issue of uncaught overstated privacy guarantees also exists for differentially-private stochastic gradient descent (DPSGD) \cite{nasr2023tight} \footnote{Our code is available at \cite{appendix}.}.

To sum up, current blackbox auditing for DP is caught in an inherent dilemma between applicability and reliability. As its primary advantage, wide applicability comes at the cost of inevitable false positives. Attempts to reduce FPs may require compromising the blackbox principle thus harming its applicability. We hope to warn data owners of approaching existing blackbox auditors with caution, and to inspire future DP auditors to strive towards resolving the current dilemma.

\begin{table}[t]
	\centering
	\caption{Presence of FPs and FNs in real-world DP mechanisms. Each FP represents a successful curator attack that compromises the data owners' privacy.}
	\label{tab:organization}
	\resizebox{1\linewidth}{!}{
		\begin{tabular}
			{p{0.18\linewidth}|p{0.2\linewidth}|p{0.2\linewidth}|p{0.21\linewidth}|p{0.17\linewidth}}
			\toprule
			\textbf{Auditors} & \textbf{DP-Sniper} \cite{bichsel2021dp} & \textbf{MPL} \cite{askin2022statistical}   & \textbf{Delta-Siege} \cite{deltasiege} & \textbf{DPSGD-Audit} \cite{nasr2023tight}               \\ \midrule
			Guarantee                 & \multicolumn{2}{c|}{$\epsilon$-DP}                 & \multicolumn{2}{c}{$(\epsilon,\delta)$-DP}                                \\ \midrule
			\multirow{3}{\linewidth}{Audited DP Mechanism} & Laplace                       & Laplace   & \multirow{3}{\linewidth}{Laplace 
				Gaussian 
			}    
			& \multirow{3}{\linewidth}{DP-SGD } \\
			& SVT & SVT   & {}    &                         \\
			& RAPPOR                    & RAPPOR   & \multicolumn{1}{c|}{}            &                         \\ \midrule
			Errors                      & FP                             & FP      & {FP $\&$ FN}       & FP                      \\ \bottomrule
		\end{tabular}
	}
\end{table}

%% file: 3_problem_and_threat.tex
\section{Preliminaries}
\subsection{Differential Privacy Auditing}
\label{sec:dp_audit}
\begin{figure}[t]
	\centering
	\includegraphics*[width=0.8\linewidth]{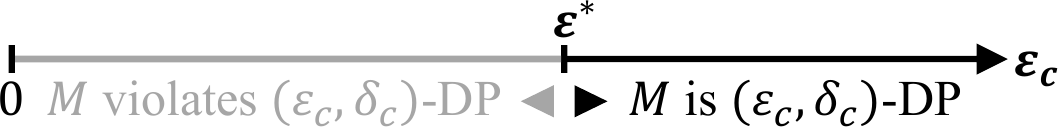}
	\caption{Illustration of the mechanism's strongest achievable DP guarantee $\epsilon^*$ and the its privacy claim $\epsilon_c$.}
	\label{fig:eps_star}
\end{figure}

\noindent\textbf{Differential Privacy.}
DP is a principled formulation of privacy-preserving mechanisms to prevent privacy leakage in data analysis. 
It requires that the probability ratio of the mechanism generating the same output on two arbitrary adjacent datasets is bounded by $e^{\epsilon}$.
Formally, a mechanism $M: \mathbb{A}\rightarrow \mathbb{B}$ is $(\epsilon_c,\delta_c)$-differentially private if for any pair of adjacent datasets $(a,a')\in\mathbb{A}^2$ and all outcome sets $S \subset \mathbb{B}$,
$
\Pr[M(a) \in {S}] \leq e^{\epsilon_c}   \Pr[M(a') \in {S}]+\delta_c.
$
\begin{definition} [True Privacy Level $\epsilon^*$]
	Given $\delta_c$, the strongest achievable DP guarantee of a mechanism $M$ is defined by
	\begin{equation*}
		\epsilon^*:= \min\{\epsilon| \forall (a,a',S), \Pr[M(a)\in S]\leq e^{\epsilon}\Pr[M(a')\in S]+\delta_c\}.
	\end{equation*}
	\label{def:eps_star} 
\end{definition}
A smaller $\epsilon^*$ indicates stronger privacy. Hence $M$ cannot achieve any $(\epsilon_c,\delta_c)$-DP where $\epsilon_c<\epsilon^*$, as illustrated in Figure \ref{fig:eps_star}.

\textbf{DP Auditing.} 
Consider the realistic scenario where a data collector releases a mechanism $M$ that claims to satisfy $(\epsilon_c, \delta_c)$-DP.
A crucial question  arises:
\begin{center}  
	\emph{Does $M$ actually afford $(\epsilon_c,\delta_c)$-DP as claimed?} 
\end{center}
A series of auditing tools are devised to validate the privacy claim $\epsilon_c$ if possible and discard it where not \cite{bichsel2018dp,ding2018detecting,wang2020checkdp,askin2022statistical,bichsel2021dp,deltasiege}. 
Auditing tools with various access assumptions approach this problem by computing the \emph{power}
\begin{equation*}
	\xi(a,a',S,\delta_c):=|\ln(\Pr[M(a)\in S]-\delta_c)- \ln(\Pr[M(a')\in S])|
\end{equation*}
of a certain \emph{witness} $(a,a',S)$, given the claimed parameter $\delta_c$. 
The power suggests to what extent the mechanism output is distinguishable.
If an auditing tool finds a witness with power exceeding $\epsilon_c$,  the claimed $(\epsilon_c,\delta_c)$-DP does not hold. The \textit{maximal power} is obtained on the \emph{theoretical optimal outcome set} $S^*$, defined as
\begin{equation}
	\forall S\subset \mathbb{B},\ \xi(a,a',S,\delta_c) \leq  \xi(a,a',S^*, \delta_c).
	\label{eq:def_optimal_S}
\end{equation}
The auditing schemes can typically identify the optimal adjacent $(a,a')$ using established pattern heuristics \cite{ding2018detecting}. However, they are not guaranteed to pinpoint $S^*$, and it is impossible to enumerate all permissible witnesses and check their corresponding power individually. 
Hence various auditing algorithms \cite{nasr2021adversary,askin2022statistical,wang2020checkdp,yang2021differential} are devised to locate a triple $(a,a',\hat{S})$ that approximately maximizes the power, referred to as the auditing tool's \emph{empirical witness}.
\begin{definition}[Maximal Power $\xi^*$]
	\label{def:maximal power} Given $\delta_c$, an auditing tool's maximal attainable power against a mechanism $M$ is defined by   
	$
	\xi^*:=\xi(a,a',\hat{S},\delta_c).
	$
\end{definition}
The auditing scheme either discards or confirms the privacy claim by comparing $\xi^*$ with $\epsilon_c$.
If $\xi^*>\epsilon_c$, the tool asserts that $M$ violates $\epsilon_c$-DP, suggesting either faulty mechanism designs or incorrect code implementations. Otherwise, the privacy claim is deemed valid as no $\epsilon_c$-DP violation is detected. 

It further follows from Def \ref{def:eps_star} that $\epsilon^*=\xi(a,a',S^*)$.
Therefore, as marked blue in Figure \ref{fig:eps_star}, the  $\xi^*$ attained by an auditing tool should always lower bound the theoretical $\epsilon^*$. The goal of auditing is to approach $\epsilon^*$  as closely as possible.

\definecolor{darkyellow}{RGB}{220, 120, 0}
\begin{table*}[h]
	\renewcommand{\arraystretch}{1.25}
	\caption{Detailed procedure of existing sampling-based blackbox auditing tools. Operations of ignoring small probabilities (DP-Sniper \cite{bichsel2021dp}) and ignoring small densities (MPL \cite{askin2022statistical}) are marked yellow.}
	\label{tab:tools}
	\resizebox{\linewidth}{!}{
		\begin{tabular}{p{0.08\linewidth}|p{0.22\linewidth}|p{0.189\linewidth}|p{0.33\linewidth}|p{0.202\linewidth}}
			\toprule
			& \textbf{Step 1}: Collect samples  
			& \textbf{Step 2}: Estimate likelihood 
			& \textbf{Step 3}: Construct empirical outcome set $\hat{S}$ & \textbf{Step 4}: Compute $\xi^*$ \\
			\midrule
			\textbf{MPL} \cite{askin2022statistical}
			& \multirow{8}{\linewidth}{Query mechanism $M$ with $a$ and $a'$ for $N$ times respectively, collect the output samples $b=M(a)$ and $b=M(a')$.}
			& Use KDE to estimate densities $p[b|a]$ and $p[b|a']$.
			& $\hat{S}=\hat{b}=\arg\max_b \left|\dfrac{p^{\textcolor{darkyellow}{\geq \tau}}(b|a)}{p^{\textcolor{darkyellow}{\geq \tau}}(b|a')}\right|$
			& $\xi^*=\left|\ln \left( \dfrac{p^{\textcolor{darkyellow}{\geq \tau}}(\hat{S}|a)}{p^{\textcolor{darkyellow}{\geq \tau}}(\hat{S}|a')} \right)\right|$
			\\
			\cline{1-1}  \cline{3-5}
			\multirow{2}{\linewidth}{\textbf{DP-Sniper} \cite{bichsel2021dp}}  
			& 
			& \multirow{6}{\linewidth}{Train classifer to estimate the likelihood ratio $r(b)$.}
			&  $\Pr[b\in \hat{S}]= 
			\begin{cases}
				1\quad \text{if}\ r(b)>\hat{t}, \\
				q\quad \text{if}\ r(b)=\hat{t}.
			\end{cases}$ 
			&\multirow{2}{\linewidth}{$\xi^*=\ln\left(\dfrac{\Pr^{\textcolor{darkyellow}{\geq c}}[M(a)\in \hat{S}]}{\Pr^{\textcolor{darkyellow}{\geq c}}[M(a')\in \hat{S}]}\right)$}   
			\\
			&
			& 
			& $(t,q)$ chosen $s.t.$ $\textcolor{darkyellow}{\Pr[M(a')\in \hat{S}]=c}$.
			& 
			\\
			\cline{1-1}  \cline{4-5}
			\multirow{4}{\linewidth}{\textbf{Delta-Siege} \cite{deltasiege}}
			& 
			& 
			&  $\hat{S}= \{b| r(b)>t\}$. $t$ chosen to minimize
			& \multirow{4}{\linewidth}{Given claimed $\delta_c$, solve $\rho(\xi^*,\delta_c)=\rho^*$.}
			\\
			&
			&
			& $\rho^*=\mathop{\min}\limits_{\epsilon,\delta}\left\{
			\rho(\epsilon,\delta)\bigg|
			\dfrac{\Pr[M(a)\in \hat{S}]-\delta}{\Pr[M(a')\in \hat{S}]} \geq e^\epsilon 
			\right\}$, where $\rho(\epsilon,\delta)$ is any function non-increasing in both $\epsilon$ and $\delta$.
			\\
			\hline 
			\multirow{3}{\linewidth}{\textbf{DPSGD-Audit} \cite{nasr2023tight}} &  \multirow{3}{\linewidth}{Run one-step DPSGD $N$ times on $a$ and $a'$, compute samples $b$ according to Alg 2 in \cite{nasr2023tight}.} 
			& Estimate the likelihood ratio $r(b)$. & $\hat{S}= \{b| r(b)>t\}$. $t$ chosen to maximize $\xi(\hat{S})=$
			$\!\max\!\left\{ \ln((
			{\Pr}[M(a)\in \hat{S}]\!-\!\delta_c)/
			\Pr[M(a')\in \hat{S}]),\right.$
			&  $\xi^*=\xi(\hat{S})$.
			\\
			&  &  &  $\left. \! \ln(
			1\!-\!{\Pr}[M(a')\!\in \!\hat{S}]\!-\!\delta_c)/
			(1\!-\!\Pr[M(a)\!\in\! \hat{S}])\!
			\right\}\!$. 
			&
			\\ \bottomrule
		\end{tabular}
	}
\end{table*}
\subsection{Sampling-Based Blackbox Auditing for DP}\label{sec:sample}
\textbf{Auditing tool's assumption.} A range of auditors has access to the curator mechanism's inner structure to facilitate determining the optimal witness. In the absence of the information, sampling-based blackbox auditing tools come into play. They are free to choose the adjacent datasets $a$ and $a'$, and collect the corresponding mechanism's output samples, but they have no knowledge of the mechanism's design or code implementation. Relying solely on mechanism outputs,
blackbox auditing is free from most scenario constraints, and is thus adaptable to a wide range of auditing user cases.

\textbf{User cases of blackbox auditing.}
To facilitate understanding, we discuss two concrete scenarios from \cite{askin2022statistical}: 1) When skeptical users request third parties to audit a mechanism, but the data collector is reluctant to reveal its proprietary mechanism design and code implementation. Blackbox assumption safeguards intellectual property while enabling effective auditing. 2) When the mechanism under audit is so complex that considering its design is infeasible, for example involving intricate hash functions as in RAPPOR \cite{rappor} or multiple compositions as in SVT \cite{svt}. In these cases, auditing schemes that require additional access are either impossible or overly cumbersome to implement, while blackbox tools remain easy to deploy.

However, due to their limited queries and absence of knowledge about the inner workings of the mechanism $M$, blackbox auditing schemes face inherent limitations:  the distributions of $M(a)$ and $M(a')$  are not analytically known. To compute $\xi^*$, they have to make nonparametric estimations about the probabilities from collected samples instead, which inevitably induces estimation errors. Below we introduce the common intuitions of representative blackbox auditing schemes and then formalize how they handle this limitation respectively.

\textbf{Representative blackbox auditing schemes.} DP-Sniper \cite{bichsel2021dp} and MPL \cite{askin2022statistical} target only the $\epsilon$-DP guarantee, whereas Delta-Siege audit the general $(\epsilon, \delta)$-DP mechanisms. We present a sketch of their algorithms in Table~\ref{tab:tools}. First, the auditor queries the targeted mechanism on adjacent datasets multiple times to collect the output samples. Then the auditor computes the likelihood ratio of the output samples:
\begin{equation}
	r(b):=\dfrac{p(b|a)}{p(b|a')},
\end{equation}
where $p(b|a)$ and $p(b|a')$ are probability density (or mass) functions for continuous (or discrete) outputs. Third, to achieve high power, the outcome set should pick outputs that are most likely to originate from $a$ rather than $a'$, i.e.,
$b$ with a large $r(b)$. A simple exercise shows that for $\epsilon$-DP, the theoretical optimal outcome set $S^*$ defined in Eq. \eqref{eq:def_optimal_S} consists only of outputs with the largest ratio. A detailed analysis is provided in Appendix A \cite{appendix}. Formally,
\begin{equation}
	S^*=\left\{
	\begin{aligned}
		& \{b^*| \forall b\in \mathbb{B}, r(b)\leq r(b^*)\} \ \text{for} \ \delta_c=0; \\
		& \arg\mathop{\max}\limits_{S}\left\{\ln
		\tfrac{\Pr[M(a)\in S]-\delta_c}{\Pr[M(a')\in S]},
		\ln
		\tfrac{\Pr[M(a)\in S]-\delta_c}{\Pr[M(a')\in S]}
		\right\} \ \text{for} \ \delta_c \not=0.
	\end{aligned}
	\right.
	\label{eq:optimal_outcome_ratio}
\end{equation}
For the relaxed $(\epsilon,\delta)$-DP, 
the parameter $\delta$ suggests that the outcome set achieving the largest $\xi$ does not strictly have to be the smallest possible set. However, since the magnitude of $\delta$ is typically very low, the optimal $S^*$ most likely remains within the small probability region.

\textbf{Omitting small probabilities (densities).}
As the auditor moves toward a smaller $S$, the probabilities $\Pr[M(a)\in S]$ and $\Pr[M(a')\in S]$ decrease, and the difficulties in assessing these small probabilities emerge. This is because given limited queries, an adversary is almost impossible to observe output samples in the small probability regime. Even if such query outputs are observed, estimating small probabilities from samples introduces large variance to the result, making the estimation highly unreliable. 

To avoid errors from unreliable estimation, DP-Sniper proactively ignores small probability outcomes altogether. Specifically, it picks a pre-set threshold $c$ and uses 
\begin{equation*}
	{\Pr}^{\geq c}[M(A)\in S]:=\max\{\Pr[M(A)\in S],c\}
\end{equation*}
in place of $\Pr[M(A)\in S]$. Only large probabilities above $c$ are used as is, thus eliminating the effect of small probabilities when computing $\xi^*$. The optimal outcome set no longer reduces to a single point, but contains outputs with the largest ratio and satisfies $\Pr[M(a')\in \hat{S}]=c$. To ensure its existence, $\hat{S}$ in DP-Sniper is randomized, i.e., it includes each outcome with a probability $q\in[0,1]$. The detailed auditing procedure is in Table \ref{tab:tools} \footnote{We use ratio estimation instead of the original posterior estimation in \cite{bichsel2021dp}.
	This is only a representational adjustment to align with MPL. It does not change the underlying technical procedure. By definition, the posterior is $p(a|b) =p(b|a)/(p(b|a)+p(b|a'))$, assuming the prior is $p(a)=p(a)'=1/2$. Then $p(a|b) = 1/(1+1/r(b))$, and hence the posterior and the ratio are equivalent.}.

MPL \cite{askin2022statistical} uses kernel density estimation to directly estimate the densities per single output point.
$\xi^*$ is unstable when the true density is close to 0 because small errors in the density estimation will translate into great errors in the logarithm. Therefore, similar to DP-Sniper, MPL \cite{askin2022statistical} ignores small densities by using a pre-set threshold $\tau$ to truncate its estimated density $p^{\geq \tau}(b|A):=\max\{p(b|A),\tau\}$ for any output $b$ and dataset $A$. It selects its $\hat{S}$ as the individual output that maximizes the truncated likelihood ratio.

Delta-Siege \cite{deltasiege}  also prioritizes output samples with a large ratio $r(b)$. It does not voluntarily cut off small probabilities, yet it still hardly observes small probability outputs. It further introduces a privacy surrogate $\rho(\epsilon,\delta)$, 
which can be any non-increasing function in both $\epsilon$ and $\delta$. All $(\epsilon,\delta)$ guarantees sharing the same $\rho$ value are considered equivalent, and a smaller $\rho$ signifies weaker privacy.  
The auditor then varies its ratio threshold $t$ to obtain multiple pairs $\Pr[M(a)\in \hat{S}]$ and $\Pr[M(a')\in \hat{S}]$. Each probability pair pinpoints a set of infeasible $(\epsilon,\delta)$-DP guarantees, from which the auditor selects the particular $(\epsilon,\delta)$ with the smallest $\rho$. Finally, the auditor chooses the smallest $\rho$ across all probability pairs as its ultimate privacy evaluation, and computes the distinguishability $\xi^*$ at the specified $\delta_c$ accordingly.

DPSGD-Audit is mostly similar to DP-Sniper. The only differences are that the collected samples are DPSGD outputs instead of statistical query outputs, and that its search for the optimal outcome set $S^*$ involves non-zero $\delta_c$ as in Eq. \eqref{eq:optimal_outcome_ratio}.

To sum up, with the truncated probabilities, the auditing tool's maximal power in the final step is unified by
\begin{equation}
	\resizebox{0.92\hsize}{!}{$
		\!\!\!\xi^* \!= \!
		\left\{
		\begin{aligned}
			&\ln(
			{\Pr}^{{\geq c}}[M(a)\in \hat{S}]
			/
			{\Pr}^{{\geq c}}[M(a')\in \hat{S}]
			), \text{ DP-Sniper};\\
			&|\ln(p^{\geq \tau}(\hat{S}|a)/ p^{\geq \tau}(\hat{S}|a'))|, \text{ MPL};\\
			& \epsilon \ \text{s.t.} \ \rho(\epsilon,\delta_c)=\rho^*,\ \text{Delta-Siege}; \\
			&\!\max\!\left\{ \ln\!\tfrac{
				{\Pr}[M(a)\in \hat{S}]-\delta_c}
			{\Pr[M(a')\in \hat{S}]},
			\ln\!\tfrac{
				1-{\Pr}[M(a')\in \hat{S}]-\delta_c}
			{1-\Pr[M(a)\in \hat{S}]}\!
			\right\}\!, \text{DPSGD-Audit}.
		\end{aligned}
		\right.
		\label{eq:xi}
		$}
\end{equation}

%% file: 4_motivation.tex
\begin{figure*}[t]
	\centering
	\includegraphics[width=0.95\linewidth]{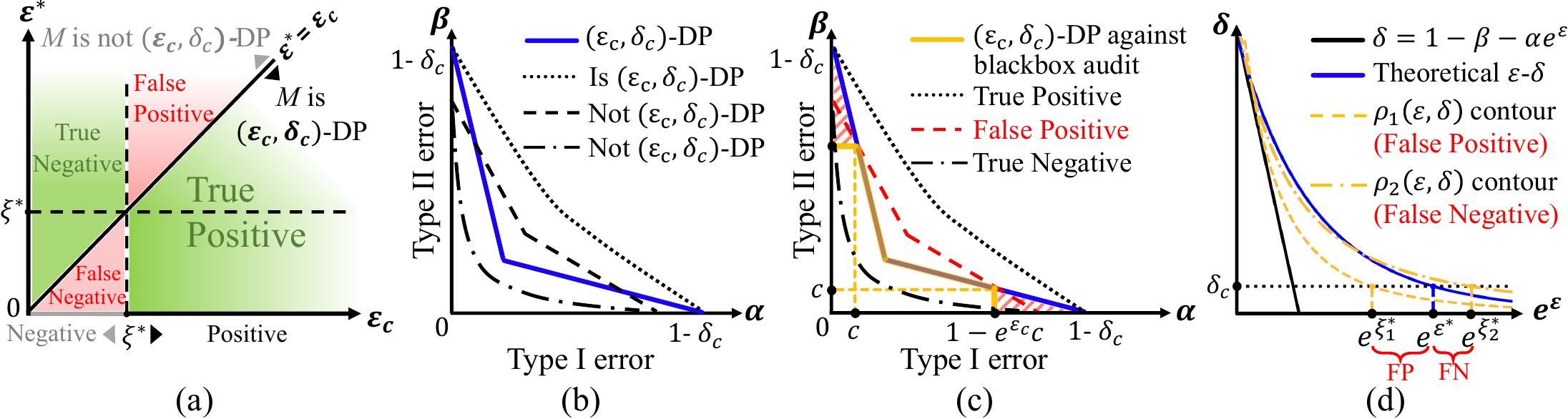}
	\caption{(a) An illustration of the FP, FN, TP, and TN regions divided by the true privacy level $\epsilon^{*}$, the claimed level $\epsilon_c$, and maximal power $\xi^*$. (b)(c) DP vs. DP against blackbox audit. FPs exist since the audit ignores small probabilities (densities). (d) Delta-Siege uses a privacy surrogate $\rho$ to seek the optimal power $\xi^*$. Different $\rho$s lead to FPs or FNs depending on the position with the theoretical $\epsilon-\delta$ DP curve. The illustrative privacy surrogates are $\rho_1(\epsilon,\delta)=e^{-3\epsilon}/\delta$ and $\rho_2(\epsilon,\delta)=e^{-2\epsilon}/\delta$.
	}
	\label{fig:tradeoff}
\end{figure*}

\section{Threat Model and Formulation}
\label{sec:motivation}
\textbf{Curator attack}. 
In data analysis scenarios, the data owner holds the raw data while the curator analyzes it and releases the outcomes. To comply with privacy regulations, the data is typically sanitized in one of two means: either the data owner applies DP mechanisms provided by the curator and sends the sanitized data to the curator for analysis, or the curator collects the raw data and applies DP algorithms themselves. In both cases, the curator is responsible for devising a sanitization mechanism that meets the required privacy level. Meanwhile, the data owner relies on blackbox auditors to inspect if the provided mechanism truly achieves its privacy claim.

We assume a malicious curator who runs DP mechanisms not conforming to its privacy claim, while escaping the auditors' detection by exploiting the loopholes in the auditing tool.
It directly compromises the privacy of data owners, hence termed a \emph{curator attack}.
Clearly, the attack is only feasible due to the inadequacy of the auditor. Therefore, it is essential to get an in-depth understanding of the current auditing schemes and be aware of the conditions under which they fail.

In the upcoming sections, we will explore the criteria for effective auditing through the lens of hypothesis testing. We will show that prior evaluations of blackbox auditors are incomplete and fail to reveal significant vulnerabilities within the tools, namely the false positive (FP) errors in hypothesis testing. Each FP constitutes a successful curator attack, as elaborated later in this section.
This oversight occurs because these existing tools wrongly consider ignoring small probabilities harmless, as demonstrated against some classical well-constructed and flawed DP mechanisms.

We emphasize that this risk is not just theoretical; it is prevalent in real scenarios. Curator attacks almost always happen whenever the data owner resorts to an existing blackbox auditor, as malicious curators are \emph{both tempted and able} to deploy an FP mechanism.
Specifically, a curator has two incentives: (1) to divulge the owners' private data; or (2) to reduce privacy protection in exchange for better data utility. It is able to do so by using an existing FP or, if none exists, by tweaking the mechanism to craft a new FP, as detailed in \S\ref{sec:roadmap}. Either way, the curator is able to deploy a false positive mechanism that fulfills the attack.
We then instantiate the construction of the attack with concrete real-life mechanisms and auditors in \S\ref{sec:fp_sniper}-\ref{sec:fp_sgdaudit}.
This real-world implication indicates that the efficacy of blackbox auditing is questionable and should be reassessed.


\textbf{A hypothesis test.} We formally state our problem of auditing the blackbox auditing tool as a hypothesis testing problem $\mathcal{P}_1$ where the auditing target is mechanism $M$ with privacy claim $\epsilon_c$:
\begin{equation*}
	\begin{aligned}
		&H_0:  \ M\ \text{is}\ (\epsilon_c,\delta_c)\text{-DP}. \\
		& H_1:   \ M\ \text{violates}\ (\epsilon_c,\delta_c)\text{-DP}.
	\end{aligned}
\end{equation*}
Confirming or discarding the privacy claim based on $\xi^*$ in \S\ref{sec:dp_audit} corresponds to retaining or rejecting the null hypothesis $H_0$, i.e., if $\xi^*\leq \epsilon_c$, we retain $H_0$ which is the positive case; otherwise, we reject $H_0$ which is the negative case. The auditing performance consists of four outcomes: true positive (TP), true negative (TN), false positive (FP) and false negative (FN). We illustrate their definition and practical interpretations in Figure \ref{fig:tradeoff}(a). 

Each mechanism $M$ under scrutiny corresponds to a distinct point $(\epsilon_c, \epsilon^*)$ in the two-dimensional plane of Figure \ref{fig:tradeoff}(a). It is also associated with a unique auditing power $\xi^*$ against a designated auditing tool. Following from Eq. \eqref{def:eps_star}, whether $M$  is in fact $\epsilon_c$-DP hinges on the comparison between $\epsilon^*$ and $\epsilon_c$, and is thus separated by the solid diagonal line in Figure \ref{fig:tradeoff}(a).
Meanwhile, the audit's assertation of ``Positive" or ``Negative'' rests upon $\xi^*$ versus $\epsilon_c$, hence the vertical dashed line. This division segments the entire plane into four regions, representing TP, TN, FP, and FN respectively. 

\textbf{A fresh look at FP.} Prior evaluations typically focus on the TP and TN regions while neglecting the FP analysis: for mechanisms that are correctly implemented, the auditing tools confirm the privacy claim with $\xi^*$ being a nearly tight lower bound of $\epsilon_c$ (TP); for mechanisms that are wrongly devised, the auditing tools report a $\xi^*$ that significantly exceeds $\epsilon_c$ (TN).
FP was mostly overlooked: few addresses the case where a mechanism is wrongly devised $(\epsilon^* > \epsilon_c)$ without the auditor catching it $(\xi^* \leq \epsilon_c)$.
Specifically, DPSGD-Audit \cite{nasr2023tight} mentions the intuition behind the blackbox auditor's limitation with a toy example, but is confined to the DPSGD mechanism and lacks a systematic quantitative analysis; MPL \cite{askin2022statistical} briefly addresses the limitation but only with one special case, which downplays the problem's severity; DP-Sniper \cite{bichsel2021dp} and Delta-Siege \cite{deltasiege} do not mention it at all.

However, FP analysis is critical because a false positive directly results in a successful curator attack by definition.
More importantly, we point out that FPs always exist since blackbox auditing schemes cannot handle small probabilities (densities). We will use the canonical $f$-DP formulation \cite{dong2019gaussian} to explain the point. By \cite{dong2019gaussian}, DP can be completely characterized by the type I error $\alpha$ and type II error $\beta$ of the following hypothesis testing $\mathcal{P}_2$:
\begin{equation*}
	\begin{aligned}
		&H_0^{\text{DP}}:\ \text{the input dataset is}\ a'.\\
		&H_1^{\text{DP}}:\  \text{the input dataset is}\ a.
	\end{aligned}
	\label{eq:dp_test}
\end{equation*}
Note that the testing problem $\mathcal{P}_2$ is different from $\mathcal{P}_1$. $\mathcal{P}_1$ assesses whether an auditor adheres to its privacy claim with a focus on evaluating the auditing capability: a more powerful auditing tool should distinguish $H_0$ from $H_1$ with a higher accuracy. In contrast, $\mathcal{P}_2$ describes how well a mechanism protects individual data records: being harder to differentiate $H_0^{\text{DP}}$ and $H_1^{\text{DP}}$ means a higher level of privacy of the mechanism. 

As visualized in Figure \ref{fig:tradeoff}(b), $(\epsilon_c,\delta_c)$-DP is parametrized by the following tradeoff of $\mathcal{P}_2$ (the blue solid curve) \cite{dong2019gaussian}
\begin{equation*}
	\inf_{\phi}\{\beta_{\phi}:\alpha_{\phi}\leq \alpha\}= \max\{0, 1-\delta_c-e^{\epsilon_c}\alpha, e^{-\epsilon_c}(1-\delta_c-\alpha)\}
\end{equation*}
for $\alpha\in [0,1]$. $S$ represents the rejection region for $H_0^{\text{DP}}$, so the type I  error is $\alpha=\Pr[M(a')\in S]$ and the type II error is $ \beta=1-\Pr[M(a)\in S]$. A mechanism is $(\epsilon_c,\delta_c)$-DP only if its entire tradeoff curve lies above the blue line.
However, when considering DP-Sniper which ignores probabilities below threshold $c$, satisfying $(\epsilon_c,\delta_c)$-DP against blackbox auditing becomes looser:
\begin{equation*}
	\forall (a,a',S),\ {\Pr}^{\geq c}[M(a) \in {S}] \leq e^{\epsilon_c}   {\Pr}^{\geq c}[M(a') \in {S}]+\delta_c.
\end{equation*} 
This transformation results in a \emph{pseudo tradeoff curve}, represented by the yellow solid line in Figure \ref{fig:tradeoff}(c). The curve is unaltered on $\alpha\in [c,1-\delta_c-e^{\epsilon_c}c]$ while requiring
\begin{equation*}
	\begin{aligned}
		\beta&
		\geq 1-\delta_c-e^{\epsilon_c}{\Pr}^{\geq c}[M(a')\in S]=1-\delta_c-e^{\epsilon_c}c,
	\end{aligned}
\end{equation*}
if $\alpha$ falls below $c$, indicated by the horizontal yellow line segment. The symmetric holds true where $\beta\leq c$. This difference in the small probability regime (the red-shaded region in Figure \ref{fig:tradeoff}(c)) allows an infinite number of mechanisms to pass the audit without actually satisfying $(\epsilon_c, \delta_c)$-DP. 
MPL's auditing \cite{askin2022statistical} follows a similar paradigm, except that it permits DP violations within small densities rather than probabilities. 

Delta-Siege is more complicated as it relies on the specific choice of the surrogate function $\rho(\epsilon,\delta)$. For the general case of auditing $(\epsilon,\delta)$-DP, we could write the surrogate in Table \ref{tab:tools} as 
\begin{equation}
	\rho^*=\mathop{\min}\limits_{\epsilon,\delta}\left\{
	\rho(\epsilon,\delta)\bigg|
	\dfrac{1-\beta-\delta}{\alpha} = e^\epsilon 
	\right\}.
\end{equation}
The observed samples could lead to different pairs of $(\alpha,\beta)$ and each $(\alpha,\beta)$ pair corresponds to a distinct line in the $e^\epsilon$-$\delta$ plane of Figure \ref{fig:tradeoff}(d) (marked as the black solid line), which represents the set of feasible $(\epsilon,\delta)$ values at each $(\alpha,\beta)$. Each $\rho$ value corresponds to a distinct contour marked as the yellow dashed curve, the shape of which depends on the specific function we choose, and the contour moves strictly towards the upper-right corner with a decreasing $\rho$. Hence the minimal $\rho^*$ is reached at the contour closest to the upper-right corner while intersecting the black solid line. As an example, for convex $\rho$ contours, this occurs where the contour is tangent to the solid line. The auditing result $\xi^*$ lies in the intersection of the $\rho^*$ contour and $\delta=\delta_c$. Clearly, $\xi^*=\epsilon^*$ only if the chosen $\rho$ contour is identical to the theoretical $e^\epsilon$-$\delta$ curve of $M$ (shown by blue solid line), or the chosen contour happens to intersect $\delta=\delta_c$ at exactly $\epsilon=\epsilon^*$. Both cases are virtually impossible for the blackbox auditor, as it is unaware of the mechanism details. Thus different choices of $\rho$ would lead to the discrepancies between $\xi^*$ and $\epsilon^*$, which turn out to be the FP and FN cases, as shown in Figure \ref{fig:tradeoff}(d).


With the above theoretical intuition, we mainly discuss the practical existence of FPs in existing auditing tools in the upcoming sections.

%% file: 5_roadmap.tex
\section{Roadmap for False Positive Analysis}
\label{sec:roadmap}

We begin with a formal definition of the FP in our hypothesis testing problem $\mathcal{P}_1$:
\begin{definition}[False Positives]
	\label{def:fp1}
	A mechanism $M$ with privacy claim $(\epsilon_c,\delta_c)$ is an FP against an auditing tool if 
	\begin{equation}
		\label{eq:fp_iff}
		\xi^*\leq \epsilon_c<\epsilon^*,
	\end{equation}
	i.e., $M$ violates $(\epsilon_c,\delta_c)$-DP ($\epsilon_c<\epsilon^*$) but passes the auditing ($\xi^*\leq \epsilon_c$).
\end{definition}

False negatives are defined similarly, i.e., $\xi^*>\epsilon_c\geq \epsilon^*$.
Given the above definition, our ultimate goal is to seek the feasible region of the DP parameter(s) satisfying  Eq.~\eqref{eq:fp_iff}. There are two possibilities: existing DP mechanisms with the parameters in the above range are naturally FPs, and we call them \textit{benchmark mechanisms in loose audit region}; in the case where such parameters are non-existent, we tweak DP mechanisms to make the feasible region non-empty, and we refer to them as \textit{adapted mechanisms in tight audit region}. We unify the two cases in Alg. \ref{alg:roadmap} where line 1-5 represents the first case and line 6-12 shows the second one. We denote the mechanism parameter(s) as $\vartheta$, (e.g., the scale of the added noise) and write $\epsilon^*$ and $\xi^*$  as functions of $\vartheta$. For clarity, we represent the parameter of benchmark and adapted mechanisms as $\vartheta=\theta$ and $\vartheta=\widetilde{\theta}$ respectively. Other notations are summarized in Table \ref{tab:craft_notations}.

For both the benchmark and the adapted mechanisms, FP exists only if $\xi^*(\vartheta)<\epsilon^*(\vartheta)$ holds as a prerequisite, and such a prerequisite also serves as a guideline in search of adaptations to existing DP mechanisms. Thus we formalize the procedure of seeking/constructing FPs for the auditing tools as follows, with the detailed conditions presented in the upcoming sections:
\begin{itemize}
	\item (\emph{P1-3}):  $\xi^*(\vartheta)<\epsilon^*(\vartheta)$. Prerequisite \ref{remark:1}-\ref{remark:3} are for DP-Sniper, MPL, and Delta-Siege, respectively. DPSGD-Audit shares prerequisite 1 with DP-Sniper.
	We compute the mechanism's theoretical $S^*$ following Eq. \eqref{eq:optimal_outcome_ratio}, and identify when the benchmark or adapted mechanism conforms to the prerequisites.
	\item (\emph{R1}): $\epsilon^*(\vartheta)>\epsilon_c$. We derive $\epsilon^*=\xi(a,a',{S}^*)$ and solve the inequality for $\theta$ or $\widetilde{\theta}$.
	This determines the parameter that violates the DP claim.
	\item (\emph{R2}): $\xi^*(\vartheta)\leq \epsilon_c$. We compute the auditing tool's empirical $\hat{S}$  following Table \ref{tab:tools}, derive $\xi^*$  following Eq. \eqref{eq:xi} and solve the inequality for $\theta$ or $\widetilde{\theta}$. This determines the parameter that passes the auditing.
\end{itemize}
The final false positive $\theta$ or $\widetilde{\theta}$ is the intersection of solution sets for the prerequisite, {R1} and {R2}. 

The procedure of discovering FNs is constructed similarly by simply reverting the inequalities:
\begin{itemize}
	\item[] (\emph{P3'}):  $\xi^*(\vartheta)>\epsilon^*(\vartheta)$,\ (\emph{R3}):\ $\epsilon^*(\vartheta)\leq\epsilon_c$,\ (\emph{R4}):\ $\xi^*(\vartheta)> \epsilon_c$.
\end{itemize}
By definition \ref{def:maximal power}, $\xi^*$ should never exceed $\epsilon^*$. Hence we point out that FNs occur only from the erroneous $\xi^*$ expression used by the auditor. DP-Sniper, MPL, and DPSGD-Audit are free from FNs as they adhere to the primitive $\xi^*$ definition. Discussion of false negatives is limited to Delta-Siege where improper privacy surrogate is used.




\begin{table}[t]
	\centering
	\caption{Notations used for FP analysis.}
	\label{tab:craft_notations}
	\resizebox{1\linewidth}{!}{
		\begin{tabular}{l|l}
			\toprule
			\textbf{Notation} & \textbf{Definition}\\
			\midrule
			\multirow{1}*{$M_{\vartheta}$} &  Mechanism with parameter $\vartheta$\\
			\midrule
				$\vartheta=\theta$
			&Parameter of the benchmark mechanism $M_{\theta}$,\\
			&e.g. Laplacian noise scale $\theta$ of SVT \cite{svt}.\\
				$\vartheta=\widetilde{\theta}$
			& Parameter of the adapted mechanism $M_{\widetilde{\theta}}$,\\
			&e.g. crafted noise scale $\widetilde{\theta}$ of adapted SVT.\\
			\midrule
			$S^*$ & Theoretical optimal outcome set.\\
			$\hat{S}$ & Empirical outcome set.\\
			\midrule
			\multirow{2}*{
				$\epsilon^*(\vartheta)$,  $\xi^*(\vartheta)$} & $\epsilon^*$ and $\xi^*$ as functions of mechanism parameter.\\
			& $\epsilon^*(\vartheta)=\xi(a,a',S^*)$;
			$\xi^*(\vartheta)
			\stackrel{\text{Eq. }\eqref{eq:xi}}{=}
			\xi(a,a',\hat{S})$. 
			\\
			\bottomrule
		\end{tabular}
	}
\end{table}

\begin{algorithm}[t]
	\SetAlgoLined
	\KwIn{Benchmark mechanism $M_{\theta}$, user-designated privacy level $\epsilon_c$}
	\KwOut{Mechanisms marked as FP}
	Derive $\epsilon^*(\theta)$ of $M_{\theta}$ following Def \ref{def:eps_star}\;
	Derive $\xi^*(\theta)$ of $M_{\theta}$ following Eq. \eqref{eq:xi}\;
	\If{$\xi^*(\theta)\leq \epsilon_c <\epsilon^*(\theta)$}{
		mark $M_{\theta}$ as FP\;
		\tcc{If the auditing is loose, the benchmark mechanism is an FP.}
	}
	\ElseIf{$\xi^*(\theta)=\epsilon_c=\epsilon^*(\theta)$}{
		adapt ${\theta}$ into ${\widetilde{\theta}}$ following Prerequisite \ref{remark:1} or \ref{remark:2}\;
		\tcc{If the auditing against benchmark mechanism is tight, we  adapt it to craft an FP.}
		derive $\epsilon^*(\widetilde{\theta})$ of $M_{\widetilde{\theta}}$ following Def \ref{def:eps_star}\;
		derive $\xi^*(\widetilde{\theta})$ of $M_{\widetilde{\theta}}$ following Eq. \eqref{eq:xi}\;
		solve $\widetilde{\theta}$ s.t. $\xi^*(\widetilde{\theta})\leq \epsilon_c < \epsilon^*(\widetilde{\theta})$\;
		mark $M_{\widetilde{\theta}}$ as FP.
	}    
	\caption{Sketch for FP construction against blackbox auditing tools.}
	\label{alg:roadmap}
\end{algorithm}

%% file: 6_sniper.tex
\section{False Positives Against DP-Sniper}
\label{sec:fp_sniper} 

To start with, DP-Sniper was only evaluated at one particular $\epsilon_c$ for each benchmark mechanism in \cite{bichsel2021dp}. However, its auditing effectiveness may vary across the spectrum of $\epsilon_c$. There may exist some $\vartheta$ s.t. $\xi^*(\vartheta) < \epsilon_c \leq \epsilon^*(\vartheta)$, making the benchmark mechanism itself a false positive. Hence we instantiate the prerequisite as
\begin{remark}[P1]
	\label{remark:1} Given any mechanism $M_{\vartheta}$,
	the iff condition for $\xi^*(\vartheta) < \epsilon^*(\vartheta)$  against DP-Sniper's auditing with probability threshold $c$ is
	\begin{equation}
		\Pr[M_{\vartheta}(a')\in S^*]<c,
		\label{eq:sniper_if}
	\end{equation}
	where $S^*$ is the theoretical optimal outcome set in Eq. \eqref{eq:optimal_outcome_ratio}.
	That is, $M_{\vartheta}$ must satisfy Eq. \eqref{eq:sniper_if} to produce a false positive against DP-Sniper.
\end{remark}

We reserve the detailed proof for Appendix B \cite{appendix} and here provide the intuition. 
Recall that DP-Sniper's empirical witness $\hat{S}$ prioritizes outputs with a high ratio and is determined by $\Pr[M_{\vartheta}(a')\in \hat{S}]=c$. The set $S^*$ comprises only those outputs with the highest likelihood ratio, that is $\{b^*|\forall b\in \mathbb{B}, r(b)\leq r(b^*)\}$.
Therefore,
when $\Pr[M_{\vartheta}(a')\in S^*]>c$, the empirical $\hat{S}$ contains a subset of $S^*$ exclusively, resulting in $\xi^*(\vartheta)=\epsilon^*(\vartheta)$ and eliminating the possibility of false positives.
However, 
if $\Pr[M_{\vartheta}(a')\in  S^*]<c$, 
to bridge the probability gap between $c$ and $\Pr[M_{\vartheta}(a')\in S^*]$,
the empirical $\hat{S}$ is forced to include additional outputs besides $ S^*$. 
With a lower ratio, these additional outputs dilute the power on $S^*$, leading to $\xi^*(\vartheta)<\epsilon^*(\vartheta)$. Only in this case can we solve for $\vartheta$ to satisfy the FP condition in Eq. \eqref{eq:fp_iff}.

We now follow Alg. \ref{alg:roadmap} and leverage Prerequisite \ref{remark:1} to construct FP on specific examples. 
We only consider $c\in[0,1/2)$ for DP-Sniper; otherwise, $c$ is no longer a small probability. 
For ease of illustration, we introduce each benchmark mechanism and its adapted counterpart in one pseudocode: adaptations are indicated with strike-though annotations. The example of RAPPOR is moved to Appendix E \cite{appendix} due to space limit.

\input{6_sniper_lap.tex}

\input{6_sniper_svt.tex}

%% file: 6_sniper_lap.tex
\subsection{Example 1: Laplace Mechanism}
\label{sec:lap_sniper}
\textbf{Benchmark Laplace mechanism $M^{\text{lap}}_{\theta}$} claims to be $\epsilon_c$-DP and operates as in Alg. \ref{alg:lap}, where $\text{Lap}(\Delta/\theta)$ denotes the Laplace distribution with density
\begin{equation*}
	p(\nu|\theta,\Delta)=\tfrac{\theta}{2\Delta}e^{-\frac{\theta}{\Delta}|\nu|}.
\end{equation*}
Without loss of generality, let the query output be $q(a)=0$ and $q(a')=\mu \in [0, \Delta].$

\begin{algorithm}[t]
	\SetAlgoLined
	\KwIn{Dataset $A$, query $q$, sensitivity $\Delta$, parameter $\theta$/$\widetilde{\theta}$}
	\KwOut{$b$}
	\st{$\nu\sim\text{Lap}(\Delta/\theta)$}\quad $\nu \sim \widetilde{\text{Lap}}(\Delta/\widetilde{\theta})$\;
	$b=q(A)+\nu$\;
	\caption{Benchmark/adapted Laplace mechanism}
	\label{alg:lap}
\end{algorithm}

\begin{figure}
	\includegraphics[width=\linewidth]{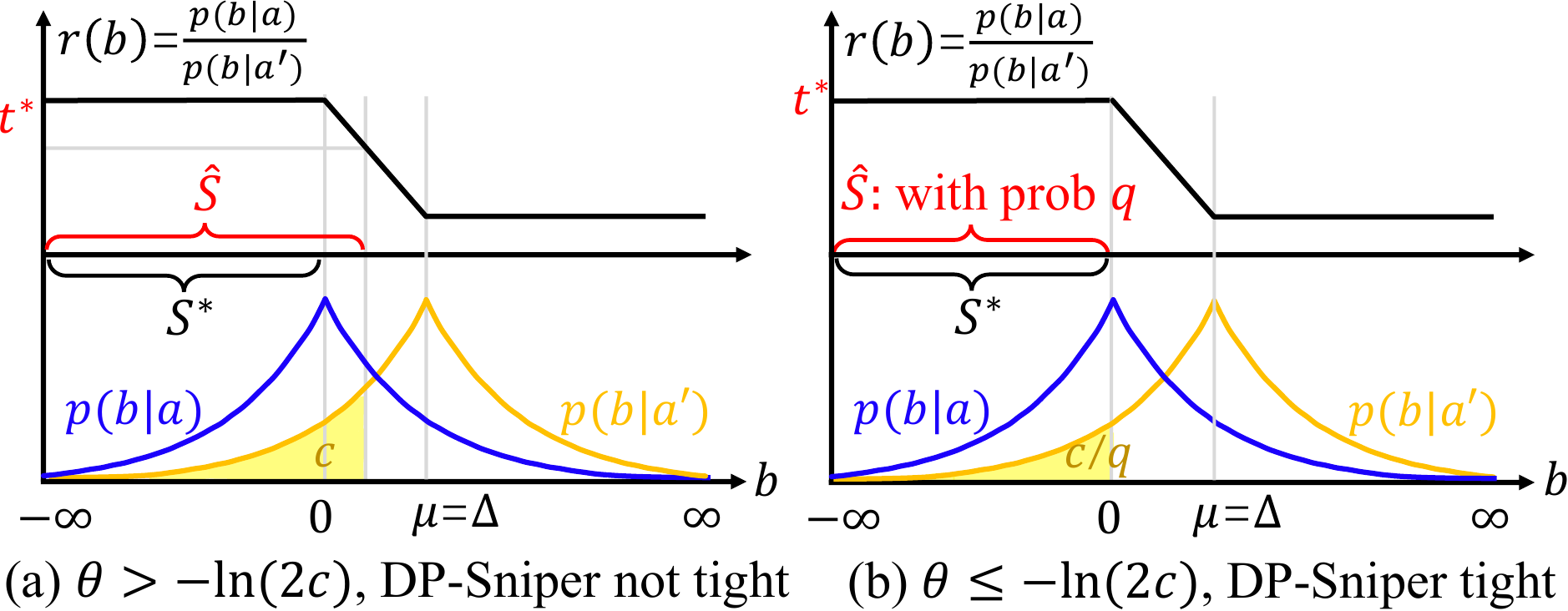}
	\caption{DP-Sniper's $\hat{S}$ against the benchmark Laplace $M^{\text{lap}}_{\theta}$. The theoretical optimal set is $S^*=(-\infty,0]$.}
	\label{fig:laplace_s}
\end{figure}

\begin{theorem}
	\label{theorem:laplace_tight}
	The benchmark Laplace mechanism $M^{\text{lap}}_{\theta}$ is an FP against DP-Sniper's auditing with probability threshold $c$ iff its privacy claim $\epsilon_c$ and parameter ${\theta}$ satisfy
	\begin{subequations}
		\begin{empheq}[left=\empheqlbrace]{align}
			& (P1) \ \ \theta>-\ln(2c), \label{eq:sniper_lap_1}\\
			&(R1) \ \  \theta >\epsilon_c,  \label{eq:sniper_lap_2}\\
			&(R2) \ \theta\leq -\ln(4c-4c^2e^{\epsilon_c}).  \label{eq:sniper_lap_3}
		\end{empheq}
		\label{eq:lap_sniper}
	\end{subequations}      
\end{theorem}
\begin{proof} 
	We show how Eq. \eqref{eq:lap_sniper} corresponds to P1, R1, and R2. Intuitively, the optimal adjacent $(a,a')$ for auditing is obtained when $q(a)=0$ and $q(a')=\Delta$ as it maximizes the difference between output distributions of $a$ and $a'$.
	
	\noindent({P1}):
	We determine the parameter $\theta$ that conforms to prerequisite \ref{remark:1}. As illustrated in Figure \ref{fig:laplace_s}, the likelihood ratio $r(b)$ is non-increasing in $b$, and the theoretical optimal outcome set $S^*$ with the largest ratio is $S^*=(-\infty,0]$. Hence $\Pr[M_{\theta}(a')\in S^*]=e^{-\theta}/2$ and prerequisite \ref{remark:1} becomes Eq. \eqref{eq:sniper_lap_1}.
	
	\noindent({R1}): We identify the parameter $\theta$ that satisfies $\epsilon^*(\theta)>\epsilon_c$.
	Following from $S^*=(-\infty,0]$, the theoretical privacy level of $M^{\text{lap}}_{\theta}$ is $\epsilon^*(\theta)=\xi(a,a',{S}^*)=\theta$. Hence R1 becomes $\theta>\epsilon_c$ as in Eq. \eqref{eq:sniper_lap_2}.
	
	\noindent({R2}): To calculate $\xi^{*}(\theta)$, we observe that DP-Sniper's empirical $\hat{S}$ selects the outputs $b$ in a descending order of $r(b)$ until the probability $\Pr[M^{\text{lap}}_{\theta}(a')\in \hat{S}]$ reaches $c$.
	The ratio is non-increasing in the output, so $\hat{S}$ is a one-sided interval of $b$; and since $c>\Pr[M^{\text{lap}}_{\theta}(a')\in (-\infty,0]]$ according to P1, the empirical $\hat{S}$ is $\Pr[b\in \hat{S}]=\mathbbm{1}\big[b\in (-\infty,(
	\tfrac{
		\ln(2c)}{\theta}
	+1)\cdot\Delta] \big]$,
	as shown in Figure \ref{fig:laplace_s}(a). Therefore, following from Eq. \eqref{eq:xi},
	$\xi^*(\theta)=\ln(
	{\Pr}^{{\geq c}}[M^{\text{lap}}_{\theta}(a)\in \hat{S}]/
	{\Pr}^{{\geq c}}[M^{\text{lap}}_{\theta}(a')\in \hat{S}])
	=\ln(1-e^{-\theta}/(4c))-\ln(c)
	$, and R2 becomes Eq. \eqref{eq:sniper_lap_3}.
\end{proof}

Alternatively, when $\theta \leq -\ln(2c)$, 
we have $\Pr[b\in S^*]>c$ and DP-Sniper's $\hat{S}$ contains the optimal $ S^*$ exclusively, as illustrated in Figure \ref{fig:laplace_s}(a).
Then $\epsilon^*(\theta)=\xi^*(\theta)$ and the benchmark Laplace mechanism no longer yield false positives.
We then adapt the noise distribution to craft a new false positive as below.

\textbf{Adapted Laplace mechanism ${M}^{\text{lap}}_{\widetilde{\theta}}$.}
Following prerequisite \ref{remark:1}, a viable adaptation should generate an outcome set where the ratio is high and the probability of $a'$ is low. Then a natural adaptation is to use a bounded noise distribution. It creates outcomes where $a'$ has zero density while $a$ maintains a non-zero density, thereby satisfying the intuition of P1. We denote such a bounded noise distribution as $\widetilde{\text{Lap}}(\Delta/\widetilde{\theta})$, with hyperparameter $\widetilde{\theta}=(\widetilde{\theta_1},\widetilde{\theta_2})$ and density
\begin{equation*}
	p(\nu|\widetilde{\theta},\Delta)=\left\{
	\begin{aligned}
		& \tfrac{\widetilde{\theta_1}}{2\Delta}e^{-\frac{\widetilde{\theta_1}}{\Delta}|\nu|},\quad |\nu| \leq \widetilde{\theta_2},\\
		& \tfrac{\widetilde{\theta_1}}{2\Delta}e^{-\frac{\widetilde{\theta_1}\widetilde{\theta_2}}{\Delta}},\quad \widetilde{\theta_2}\leq |\nu| \leq \widetilde{\theta_2}+\tfrac{\Delta}{\widetilde{\theta_1}}.
	\end{aligned}
	\right.
\end{equation*}
This particular noise distribution is not exclusive to crafting an FP. It is selected primarily for the ease of computation.
\begin{theorem}
	The adapted Laplace mechanism ${M}^{\text{lap}}_{\widetilde{\theta}}$ is an FP against DP-Sniper auditing with probability threshold $c$, if its privacy claim $\epsilon_c$ and parameter $\widetilde{\theta}=(\widetilde{\theta_1},\widetilde{\theta_2})$ satisfy
	\begin{subequations}
		\begin{align}
			&(R2) \ \
			1 \leq \widetilde{\theta_1} < \epsilon_c  \ \   \& \ \
			\frac{1}{2}e^{-\frac{\widetilde{\theta_1}\widetilde{\theta_2}}{\Delta}+\widetilde{\theta_1} - 1} \leq (e^{\epsilon_c} - e^{\widetilde{\theta_1}})c, \label{eq:subeq1} \\
			\text{or} \ \ &(R2) \ \
			\widetilde{\theta_1} < \min\{\epsilon_c, 1\} \ \   \& \ \
			\frac{\widetilde{\theta_1}}{2}e^{-\frac{\widetilde{\theta_1}\widetilde{\theta_2}}{\Delta}} \leq (e^{\epsilon_c} - e^{\widetilde{\theta_1}})c.
			\label{eq:subeq2}
		\end{align}
		\label{eq:adapted_lap_sniper}
	\end{subequations}
	\label{theorem:adapted_lap_sniper}
\end{theorem}

Please refer to Appendix C \cite{appendix} for the proof.

%% file: 6_sniper_svt.tex
\subsection{Example 2: SVT}
\label{sec:svt_sniper}
\textbf{Benchmark SVT mechanism $M^{\text{svt}}_{\theta}$.}
The SVT mechanism \cite{svt} is described in Alg. \ref{alg:altered_svt}. Compared to the previous Laplace mechanism, SVT is representative of analysis because
1) the outputs are discrete-valued vectors, and 2) the output probabilities are more complicated due to the algorithm's two-step composition. We replicate the settings from \cite{bichsel2021dp} where the abort threshold is set as $\bar{t}=1$, and the query thresholds are  $T=0.5$.

Specifically, Alg. \ref{alg:altered_svt} aborts until it outputs a symbol $\top$ or exhausts all $N$ queries, so the possible outputs are
$b^j:=[\top],\ j=0; b^j:=[\bot^{j}, \top], \ j=1, \cdots, N-1;$ and $b^j:=[\bot^{N}], j=N.$
As an example, the corresponding output probability of $b^j (j\in [1,N-1])$ is
$\Pr[M^{\text{svt}}(A)=b^j]=\int_{-\infty}^{\infty}\Pr[\rho=z]\cdot \prod_{i\in[1,j-1]}\Pr[q_i(A)+\nu_i<T_i+z]\cdot \Pr[q_{j}(A)+\nu_{j}\geq T_{j}+z]\cdot \text{d}z$ for $A=a$ or $a'$.
With the integration and product involved, SVT's output distribution is not straightforward as in \S\ref{sec:lap_sniper}.

\begin{theorem}\label{thm:benchmarkSVT}
	The benchmark SVT mechanism $M^{\text{svt}}_{\theta}$ is an FP against DP-Sniper's auditing with probability threshold $c$ if its privacy claim $\epsilon_c$ and parameter ${\theta}$ satisfy Eq. \eqref{eq:svt_sniper_benchmark}:
	\begin{subequations}
		\begin{empheq}[left=\empheqlbrace]{align}
			& (P1) \ \ \frac{2}{3}e^{-\frac{\theta}{4}}-\frac{1}{6}e^{-\frac{\theta}{2}}<c, \\
			&(R1) \ \ \ln\left(2+\frac{1}{3}e^{-\theta/2}-\frac{4}{3}e^{-\theta/4}\right) >\epsilon_c,  \\
			&(R2) \ \ln\left(\frac{c}{2}(1+\frac{c+\frac{1}{6}e^{-\theta/2}-\frac{2}{3}e^{-\theta/4}}{1+\frac{1}{6}e^{-\theta/2}-\frac{2}{3}e^{-\theta/4}})\right)\leq \epsilon_c.  
		\end{empheq}
		\label{eq:svt_sniper_benchmark}
	\end{subequations}    
\end{theorem}
The proof is similar to Thm. \ref{theorem:laplace_tight} and is deferred to Appendix D \cite{appendix}.

\begin{algorithm}[t]
	\SetAlgoLined
	\KwIn{Dataset $A$, query set $Q$, sensitivity $\Delta$, parameter $\theta/\widetilde{\theta}$, thresholds $\textbf{T}=T_1,\cdots T_{N}$, abort threshold $\bar{t}$}
	\KwOut{$b$}
	\st{$\theta_1=\theta/2$, $\rho\sim{\text{Lap}}(\Delta/\theta_1)$}\ \ \textcolor{black}{$\rho\sim \text{Uniform}(-2\widetilde{\theta_1}, -\widetilde{\theta_1})$}\;
	count $=0$\;
	\For{each query $q_i\in Q$}{
		\st{$\theta_2=\theta/2$, $\nu_i\sim{\text{Lap}}(2\bar{t}\Delta/\theta_2)$}\ \ $\nu_i\sim\text{Lap}(\bar{t}\Delta/\widetilde{\theta_2})$\;
		\If{$q_i(A)+\nu_i\geq T_i+\rho$}{
			Output $b_i=\top$\;
			count=count$+1$, \textbf{Abort} if count$\geq \bar{t}$\;}
		\Else
		{Output $b_i=\bot$\;}
	}
	\caption{Benchmark/adapted SVT mechanism}
	\label{alg:altered_svt}
\end{algorithm}

\textbf{Adapted SVT mechanism $M^{\text{svt}}_{\widetilde{\theta}}$.}
Upon inspecting the expression for $\Pr[M^{\text{svt}}(A)=b^j]$, we can see that simplifying $\Pr[\rho=z]$ facilitates the probability calculation. Hence we modify the noise $\rho$ to follow a uniform distribution.  To align with prerequisite \ref{remark:1}, we further assign a negative value to $\rho$ so that $\Pr[q_i(A)+\nu_i<T_i+z]$ can attain small values, thereby allowing the mechanism's output probability to be small as well. W.l.o.g., in Alg. \ref{alg:altered_svt},  we choose noise $\rho$ and $\nu$ to follow $\text{Uniform}(-2\widetilde{\theta_1}, -\widetilde{\theta_1})$ and ${\text{Lap}}(1/\widetilde{\theta_2})$ respectively. These distributions are merely selected for ease of probability calculation. Hence we have
\begin{theorem}\label{thm:dpsniperSVT}
	The adapted SVT mechanism $M^{\text{svt}}_{\widetilde{\theta}}$ is an FP against DP-Sniper's auditing with probability threshold $c$, if its privacy claim $\epsilon_c$ and parameter $\widetilde{\theta}=(\widetilde{\theta_1},\widetilde{\theta_2})$ satisfy  
	\begin{subequations}
		\begin{empheq}[left=\empheqlbrace]{align}
			&(P1)\ \ 
			f(\widetilde{\theta_1},\widetilde{\theta_2})<c,\label{eq:svt_sniper_sub1}
			\\
			&(R1)\ \
			\widetilde{\theta_2}>\epsilon_c,
			\label{eq:svt_sniper_sub2}
			\\
			&(R2)\ \
			e^{\widetilde{\theta_2}}f(\widetilde{\theta_1},\widetilde{\theta_2})+
			q\cdot
			(1-e^{\widetilde{\theta_2}}f(\widetilde{\theta_1},\widetilde{\theta_2}))\leq e^{\epsilon_c}c,
			\label{eq:svt_sniper_sub3}
		\end{empheq}
		\label{eq:svt_sniper}
	\end{subequations}
	where $f(\widetilde{\theta_1}, \widetilde{\theta_2}):=
	{(e^{-\widetilde{\theta_1} \widetilde{\theta_2}}-e^{-2\widetilde{\theta_1} \widetilde{\theta_2}})}
	/
	({\widetilde{\theta_1} \widetilde{\theta_2}})$ and 
	$q:=(c-f(\widetilde{\theta_1},\widetilde{\theta_2}))/$ $(1-f(\widetilde{\theta_1},\widetilde{\theta_2}))$.
\end{theorem}
The detailed proof is deferred to Appendix D \cite{appendix}.

%% file: 7_mpl.tex
\section{False Positives Against MPL}
\label{sec:fp_mpl}
Different from DP-Sniper, MPL's audit against the benchmark Laplace and SVT has been confirmed to be consistently tight across various $\epsilon_c$ regions in \cite{askin2022statistical}. Therefore, these benchmark mechanisms do not have a loose audit region, but can be adapted to others with a loose audit region, as shown in this section.

\begin{remark}[P2]
	\label{remark:2}
	Given an adapted  mechanism $M_{\vartheta}$ against MPL's auditing with density threshold $\tau$, the iff condition for $\xi^*(\vartheta) < \epsilon^*(\vartheta)$ is
	\begin{equation}
		\label{eq:mpl_if}
		\forall b\in S^*,\ \min\{p(b|a),p(b|a')\}<\tau,
	\end{equation}
	where $S^*$ is the theoretical optimal outcome set in Eq. \eqref{eq:optimal_outcome_ratio}. Hence $M_{\vartheta}$ must first satisfy Eq. \eqref{eq:mpl_if} to produce a false positive against MPL.
\end{remark}
The proof is deferred to Appendix F \cite{appendix} and the intuition is as follows. If some output $b$ in the theoretical optimal set has both densities $p(b|a)$ and $p(b|a')$ above the density threshold,  MPL can precisely estimate these densities, thereby closely approximating the real $\epsilon^*$.
False positive arises only when all theoretical optimal outputs conform to Eq. \eqref{eq:mpl_if}, leading to substantial deviations in density estimation.
Similar to \S\ref{sec:fp_sniper}, we now instantiate Alg. \ref{alg:roadmap} and Prerequisite \ref{remark:2} on the Laplace, SVT, and One-time RAPPOR (moved to Appendix G \cite{appendix}) mechanisms.

\subsection{Example 1: Laplace Mechanism}
\label{sec:lap_mpl}
\textbf{Adapted Laplace mechanism ${M}^{\text{lap}}_{\widetilde{\theta}}$.} 
Following P2, a straightforward adaptation is to leave the original Laplace distribution unchanged for densities above $\tau$, and adapt the small densities to $\tau$. Formally, the adapted noise distribution is $\widetilde{\text{Lap}}(\Delta/\widetilde{\theta})$, with density
\begin{equation*}
	p(\nu|\widetilde{\theta},\Delta)=\left\{
	\begin{aligned}
		& \tfrac{\widetilde{\theta}}{2\Delta}e^{-\frac{\widetilde{\theta}}{\Delta}|\nu|},\quad |\nu| \leq \tfrac{\Delta}{\widetilde{\theta}}\ln\tfrac{2\Delta\tau}{\widetilde{\theta}},\\
		& \tau,\quad \tfrac{\Delta}{\widetilde{\theta}}\ln\tfrac{2\Delta\tau}{\widetilde{\theta}}\leq|\nu|\leq \tfrac{\Delta}{\widetilde{\theta}}\ln\tfrac{2\Delta\tau}{\widetilde{\theta}}+\tfrac{\Delta}{\widetilde{\theta}}.
	\end{aligned}
	\right.
\end{equation*}
W.l.o.g., let $q(a)=0$ and $q(a')=\Delta$.
\begin{theorem}
	The adapted Laplace mechanism ${M}^{\text{lap}}_{\widetilde{\theta}}$ is an FP against MPL auditing with density threshold $\tau$, iff its privacy claim $\epsilon_c$ and parameter $\theta$ satisfy
	\begin{equation}
		(R2)\ \ \widetilde{\theta}\leq \epsilon_c.
		\label{eq:mpl_lap}
	\end{equation}
\end{theorem}
\begin{proof}
	\noindent ({P2}): 
	The theoretical optimal set $S^*$ of this adapted mechanism is where either $p(b|a)=0$ or $p(b|a')=0$ and the other is not $0$, with a likelihood ratio of $\infty$.
	Hence prerequisite \ref{remark:2} is naturally satisfied for any $\widetilde{\theta}$ value.
	
	\noindent ({R1}): The theoretical privacy level of ${M}^{\text{lap}}_{\widetilde{\theta}}$ is $\epsilon^*(\widetilde{\theta})=\infty$ for any $\widetilde{\theta}$, so R1 always holds as well.
	
	\noindent ({R2}): The empirical outcome set is now $\hat{S}=\{0\}$. Hence
	$\xi^*(\widetilde{\theta})=|\ln(p^{\geq \tau}(0|a))- \ln(p^{\geq \tau}(0|a'))|=\widetilde{\theta}
	$, if substituting $q(a)=0$ and $q(a')=\Delta$ into $\xi^*(\widetilde{\theta})$, we turn the condition of R2 into Eq. \eqref{eq:mpl_lap}.
\end{proof}

\subsection{Example 2: SVT}
\label{sec:svt_mpl}
\textbf{Adapted SVT mechanism ${M}^{\text{svt}}_{\widetilde{\theta}}$.}
Following prerequisite \ref{remark:2}, the proper adaptation should be able to generate outcomes with small probabilities. The intuition is similar to that in Sec. \ref{sec:svt_sniper}, so we leverage the same noise adaptation.
\begin{theorem}
	The adapted SVT mechanism ${M}^{\text{svt}}_{\widetilde{\theta}}$ is an FP against MPL auditing with probability threshold $\tau$, if its privacy claim $\epsilon_c$ and parameter $\widetilde{\theta}=(\widetilde{\theta_1},\widetilde{\theta_2})$ satisfy
	\begin{subequations}
		\begin{empheq}[left=\empheqlbrace]{align}
			&(P1)\ \ 
			f(\widetilde{\theta_1},\widetilde{\theta_2})<\tau,\label{eq:svt_mpl_sub1}
			\\
			&(R1)\ \
			\widetilde{\theta_2}>\epsilon_c,
			\label{eq:svt_mpl_sub2}
			\\
			&(R2)\ \
			e^{\widetilde{\theta_2}}f(\widetilde{\theta_1},\widetilde{\theta_2})\leq e^{\epsilon_c}\tau,
			\label{eq:svt_mpl_sub3}
		\end{empheq}
		\label{eq:adapted_svt_mpl}
	\end{subequations}
	where $f(\widetilde{\theta_1}, \widetilde{\theta_2}):=
	{(e^{-\widetilde{\theta_1} \widetilde{\theta_2}}-e^{-2\widetilde{\theta_1} \widetilde{\theta_2}})}
	/
	({\widetilde{\theta_1} \widetilde{\theta_2}})$.
\end{theorem}
The proof of the theorem is similar to the proof of Thm.~\ref{thm:dpsniperSVT} and thus is omitted.

%% file: 8_delta-siege.tex
\section{False Positives/Negatives Against Delta-Siege}
\label{sec:fp-siege}

The prerequisite of Delta-Siege differs from DP-Sniper or MPL, as the privacy surrogate $\rho(\epsilon,\delta)$ introduces potential false negatives. A detailed proof is in Appendix H \cite{appendix}.


\begin{remark}[P3/P3']
	\label{remark:3}
	Given any  mechanism $M_{\vartheta}$ against Delta-Siege's auditing with privacy function $\rho(\epsilon,\delta)$ and smallest achievable probability $c$, $\xi^*(\vartheta) \neq \epsilon^*(\vartheta)$ iff the $\rho(\epsilon,\delta)$ contour differs from $\delta=\mathcal{T}_{\vartheta}(\epsilon)$ or $\Pr[M(a')\in S^*]<c$, where $\mathcal{T}_{\vartheta}(\epsilon)$ is the mechanism's theoretical $\delta$-$\epsilon$ curve defined as\begin{equation*}
		\mathcal{T}_{\vartheta}(\epsilon):=\min\{\delta|M_{\vartheta}\ \text{satisfies}\ (\epsilon,\delta)\text{-DP}.\}
	\end{equation*}
	and $S^*$ is the theoretical optimal set in Eq. \eqref{eq:def_optimal_S}.
\end{remark}

\subsection{Example 1: Laplace Mechanism}
\label{sec:lap_siege}
\textbf{Benchmark Laplace mechanism $M_{\theta}^{\text{lap}}$} operates as in Alg. \ref{alg:lap}. Delta-Siege specifically uses $\rho(\epsilon)=\frac{\Delta}{\epsilon}$, while we point out that any non-increasing $\rho(\epsilon)$ follows the same FP analysis as below.
\begin{theorem}
	\label{theorem:laplace_siege}
	The benchmark Laplace mechanism $M^{\text{lap}}_{\theta}$ is an FP against Delta-Siege's auditing with smallest achievable probability $c$ and any $\rho(\epsilon)$ that is non-increasing and continuous in $\epsilon$, iff its privacy claim $\epsilon_c$ and parameter ${\theta}$ satisfy Eq. \eqref{eq:sniper_lap_1}, \eqref{eq:sniper_lap_2} and \eqref{eq:sniper_lap_3}.     
\end{theorem}
\begin{proof}
	\begin{figure}[t]
		\centering
		\includegraphics[width=0.98\linewidth]{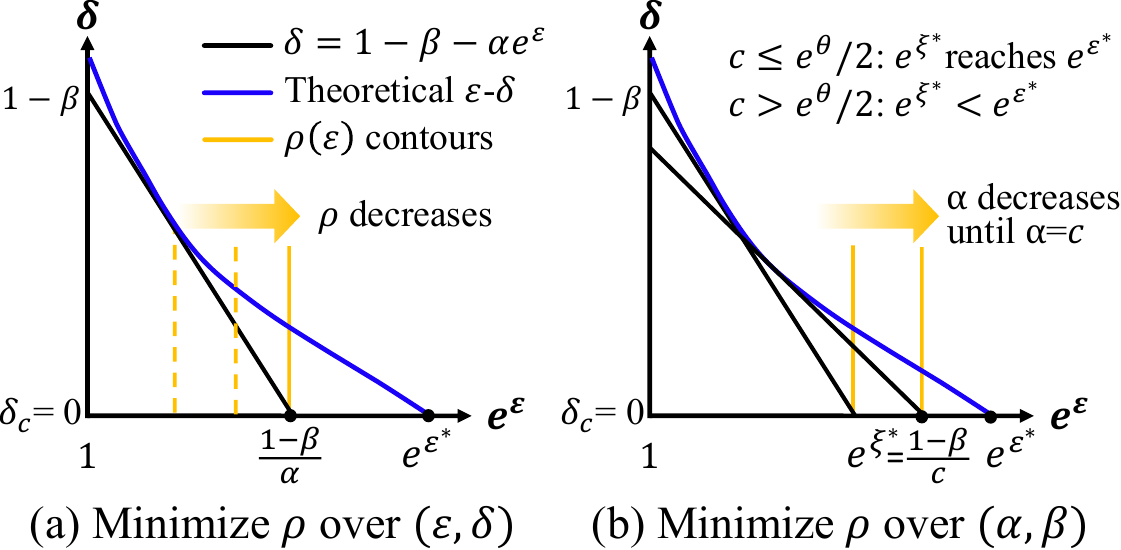}
		\caption{Laplace mechanism against Delta-Siege's auditing.}
		\label{fig:delta_siege_lap}
	\end{figure}
	We reserve the rigorous proof for Appendix  H \cite{appendix} and here visualize the intuition. 
	The theoretical $\delta-e^{\epsilon}$ curve of $M^{\text{lap}}_{\theta}$ is $\delta=\mathcal{T}(\epsilon)= 1-e^{-{\theta}/{2}}\sqrt{e^{\epsilon}}$, and the $\rho(\epsilon)$ contours are vertical lines as shown in Figure \ref{fig:delta_siege_lap}(a). 
	Given any $\alpha$ and $\beta$, the minimal 
	$\rho$ 
	is the contour closest to the upper-right corner while intersecting the line $\delta=1-\beta-\alpha e^{\epsilon}$, i.e. when $e^{\epsilon}={(1-\beta)}/{\alpha}$. 
	
	(P3):
	As auditor varies its likelihood threshold, the $(\alpha,\beta)$ pair changes while conforming to
	\begin{equation}
		\beta(\alpha)=1-e^{\theta}\alpha \ \text{if}\ \alpha < \tfrac{e^{-\theta}}{2};\ \beta(\alpha)=\tfrac{e^{-\theta}}{4\alpha}\ \text{if} \ \tfrac{e^{-\theta}}{2} \leq \alpha < \tfrac{1}{2},
		\label{eq:lap_tradeoff}
	\end{equation}
	by the definition of tradeoff function \cite{dong2019gaussian}. The pair achieves $\xi^*$ when the $\rho$ contour is rightmost, which in turn is when $\alpha$ reaches the minimum $c$, as in Figure \ref{fig:delta_siege_lap}(b).
	Therefore, the prerequisite of $\xi^*<\epsilon^*$ becomes $(1-\beta(c))/c<e^{\epsilon^*}$, which combined with Eq. \eqref{eq:lap_tradeoff} is simplified as $c \geq e^{\theta}/2$ (Eq. \eqref{eq:sniper_lap_1}). 
	
	(R1 \& R2): The rest of the proof is omitted as it is identical to DP-Sniper's Theorem \ref{theorem:laplace_tight}.
\end{proof}
We demonstrate that no FN exists for the benchmark Laplace mechanism.
When $c<e^{\theta}/2$, it follows from Eq. \eqref{eq:lap_tradeoff} that $\forall \alpha \in [c,e^{\theta}/2), (1-\beta)/\alpha=e^{\theta}$, i.e. the $e^{\xi^*}$ point in Figure \ref{fig:delta_siege_lap} coincides with the $e^{\epsilon^*}$ point. Hence $\xi^*$ never exceeds $\epsilon^*$ and FNs do not exist.

\subsection{Example 2: Gaussian Mechanism}
\label{sec:gauss_siege}
\textbf{Benchmark Gaussian Mechanism $M_{\theta}^{\mathcal{N}}$}. The benchmark Gaussian mechanism adds Gaussian noise with standard deviation $\theta$.
Below we instantiate the conditions for FP and FN when $\rho(\epsilon,\delta)=e^{-\epsilon}/\delta$. Other choices of $\rho$ follow a similar analysis.
\begin{theorem}
	Consider a Delta-Siege auditor with $\rho(\epsilon,\delta)=e^{-\epsilon}/\delta$ and a smallest achievable probability $c$.
	Define 
	\begin{equation*}\begin{aligned}
			& \beta(\alpha):=\Phi(\Phi^{-1}(1-\alpha)-\tfrac{\Delta}{\theta}),\\ 
			& S_1:=\{\alpha | c<\alpha<\tfrac{1-\beta(\alpha)}{2}\} \ \text{and} \ S_2:=\{\alpha | \tfrac{1-\beta(\alpha)}{2}\leq \alpha < 1\}.
		\end{aligned}
	\end{equation*}
	The benchmark gaussian mechanism $M_{\theta}^{\mathcal{N}}$ 
	is a false positive (false negative) against such auditor iff $\theta$ satisfies Eq. \eqref{eq:gauss_deltasiege_fp} (Eq. \eqref{eq:gauss_deltasiege_fn}).
	\begin{subequations}
		\begin{empheq}[left=]{align}
			&(FP)\ \ 
			-\ln(\delta_c\rho^*(\theta)) \mathop{\leq}\limits^{R2} \epsilon_c \mathop{<}\limits^{R1} \mathcal{T}^{-1}_{\theta}(\delta_c)
			\label{eq:gauss_deltasiege_fp}
			\\
			&(FN)\ \
			-\ln(\delta_c \rho^*(\theta))  \mathop{>}\limits^{R4}\epsilon_c \mathop{\geq}\limits^{R3} \mathcal{T}^{-1}_{\theta}(\delta_c),\label{eq:gauss_deltasiege_fn}
		\end{empheq}
		\label{eq:delta-siege_gaussian}
	\end{subequations} where $\mathcal{T}^{-1}_{\theta}(\cdot)$ is the inverse function of $\mathcal{T}_{\theta}(\epsilon)=\Phi(-\frac{\epsilon\theta}{\Delta}+\frac{\Delta}{2\theta})-e^{\epsilon}\Phi(-\frac{\epsilon\theta}{\Delta}-\frac{\Delta}{2\theta})$, and $\rho^*(\theta):=\min\{\mathop{\min}\limits_{\alpha\in S_1}\frac{1}{1-\beta(\alpha)-\alpha}, \mathop{\min}\limits_{\alpha\in S_2}\frac{4\alpha}{(1-\beta(\alpha))^2}\}$. 
	\label{theorem:delta-siege_gauss}
\end{theorem}
\begin{proof}
	\begin{figure}[t]
		\centering
		\includegraphics[width=0.95\linewidth]{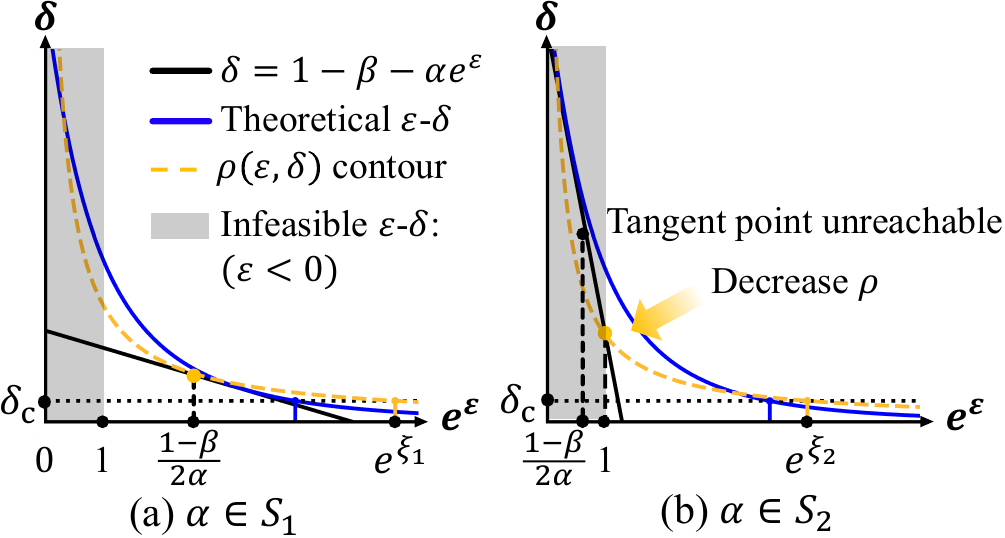}
		\caption{Gaussian mechanism against Delta-Siege's auditing.}
		\label{fig:delta_siege_gauss}
	\end{figure}
	We first compute $\xi^*(\theta)$ and $\epsilon^*(\theta)$. Then prerequisites (P3/P3') and requirements (R1-4) are directly established.
	The expression of $\beta(\alpha)$ and $\mathcal{T}_{\theta}(\epsilon)$ both follow directly from \cite{dong2019gaussian}. 
	
	Given a claimed $\delta_c$, $\epsilon^*(\theta)=\mathcal{T}^{-1}_{\theta}(\delta_c)$ by definition. The computation of $\xi^*(\theta)$ involves a two-case discussion as illustrated in Figure \ref{fig:delta_siege_gauss}. For any $(\alpha,\beta)$ pair, the convex $\rho$ contour is tangent to the line $\delta=1-\beta-\alpha e^{\epsilon}$ on the $\delta-e^{\epsilon}$ plain at $e^{\epsilon}=\frac{1-\beta}{2\alpha}$. 
	This tangent point changes across the $(\alpha,\beta(\alpha))$ pairs and may lie within the infeasible region where $\epsilon<0$ (the shaded area in Figure \ref{fig:delta_siege_gauss}). 
	Specifically, when $\frac{1-\beta}{2\alpha}>1$ (i.e. $\alpha\in S_1$), $\rho$ is minimized at the tangent point $e^{\epsilon}$, yielding $\rho=\frac{4\alpha}{(1-\beta)^2}$. When $\frac{1-\beta}{2\alpha}\leq 1$ (i.e. $\alpha\in S_2$), the tangent point is no longer valid, and the minimal $\rho$ occurs at $e^{\epsilon}=1$ instead, resulting in $\rho=\tfrac{1}{1-\beta-\alpha}$. The minimal $\rho^*$ is found across all $(\alpha,\beta(\alpha))$ pairs.
	Finally, given $\rho(\epsilon,\delta)=e^{-\epsilon}/\delta$, the auditor's computed power is $\xi^*=-\ln(\delta_c\rho^*(\theta))$.
\end{proof}

%% file: 9_dpsgd.tex
\section{False Positives in DPSGD-Audit}
\label{sec:fp_sgdaudit}
The one-step DPSGD mechanism selects two adjacent datasets, performs one step of gradient descent, and adds Gaussian noise with standard deviation $\theta$ to the gradient.
Its FP analysis closely mirrors that of DP-Sniper. The only difference is that with $\delta_c\not=0$, $S^*$ is calculated using the primitive definition in Eq. \eqref{eq:optimal_outcome_ratio} instead. Hence we defer the prerequisite and the detailed discussion to Appendix I \cite{appendix}.

\begin{theorem}
	One-step DPSGD $M^{\text{sgd}}_{\theta}$ is an FP against blackbox auditor with smallest achievable probability $c$ if its privacy claim $(\epsilon,\delta)$ and parameter $\theta$ satisfy
	\begin{equation*}
		\ln(1-\delta_c-\beta(c))-\ln(c)\leq \epsilon_c < \mathcal{T}^{-1}(\delta_c),
	\end{equation*} 
	where $\mathcal{T}^{-1}_{\theta}(\cdot)$ and $\beta(\cdot)$ are as defined in Thm. \ref{theorem:delta-siege_gauss}.
\end{theorem}

%% file: 2_related.tex
\section{Related Works}
We discuss existing approaches to auditing the DP claim of a curator mechanism. 

\textbf{Blackbox auditing for DP mechanisms.}
Blackbox auditing requires only query access to the curator's mechanism. From the collected output samples of the mechanism, it determines the optimal witness, makes nonparametric estimations on the output probabilities/densities and computes the power accordingly. The primary strength of blackbox auditing lies in its minimal access requirements, thus highly applicable to a vast array of curator mechanisms.

Representative blackbox auditing tools against DP statistical queries include DP-Sniper \cite{bichsel2021dp}, MPL \cite{askin2022statistical}, Delta-Siege \cite{deltasiege}, etc. DP-Sniper and MPL both leverage a well established heuristic to locate the optimal adjacent dataset pair \cite{ding2018detecting}. Their main focus and contribution lies in determining the optimal outcome set, which was previously intractable to solve.
DP-Sniper addresses this challenge by training classifiers to estimate the likelihood ratio and selecting the outputs with the highest ratios as the optimal outcome set. However, this method implicitly induces a parametric assumption on posterior distributions, potentially limiting DP-Sniper's effectiveness if the posterior distribution significantly deviates from the assumed class in the classifier. To improve this, MPL \cite{askin2022statistical} removes this parametric assumption by leveraging the kernel density estimator (KDE), which directly estimates the density function based on interpolation. The performance of MPL \cite{askin2022statistical} and DP-Sniper \cite{bichsel2021dp} are close, while MPL \cite{askin2022statistical} is more computationally efficient. 
Delta-Siege \cite{deltasiege} targets the more general $(\epsilon,\delta)$-DP and attempts to unify the two parameters with a privacy surrogate $\rho(\epsilon,\delta)$. It then transforms the problem of maximizing the power $\xi$ to minimizing $\rho$. However, we point out that this method is inherently flawed under the blackbox auditing setting, as blindly choosing $\rho$ without any knowledge of the audited mechanism induces numerous false positives/negatives.

Another line of work focuses on auditing the DPSGD training pipeline \cite{nasr2021adversary,zanella2023bayesian,lu2022general,jagielski2020auditing,steinke2024privacy} and deriving tighter confidence intervals of their estimates. These work typically achieve auditing via launching membership inference attacks under various attacker access. Notice that our blackbox auditing access does not suggest the auditor only has API access to the trained model. Rather, a blackbox auditor can obtain the intermediate gradients of the DPSGD training. Blackbox auditing only means the auditor does not leverage the parametric information of the added noise.

\textbf{Non-blackbox auditing for DP mechanisms.}  Compared to the auditors above, non-blackbox auditing \cite{albarghouthi2017synthesizing,wang2019proving,bichsel2018dp,ding2018detecting,ravi2019automated,wang2020checkdp,dixit2013testing,wilson2020differentially} requires knowledge of the inner structure of the curator's mechanism, such as mechanism design or code implementation. As such, As such, compared to blackbox auditing, they are more suited for mechanism developers rather than general data owners. In addition, they are not applicable to proprietary curator mechanisms and are cumbersome to deploy when dealing with complex mechanisms. 

Specifically, a line of these schemes leverage formal verification to perform DP auditing \cite{albarghouthi2017synthesizing,wang2019proving,wang2020checkdp,ravi2019automated}. Their applicability is highly limited by the symbolic solver they use. For example, CheckDP \cite{wang2020checkdp} and DiPC \cite{ravi2019automated} uses solvers that are non-compatible with hash functions. Therefore, unlike blackbox auditors, they cannot audit any curator mechanism involving hash functions, for example the RAPPOR mechanism. 

Other auditing methods make additional assumptions about the curator mechanism beyond just obtaining output samples, limiting their applicability to specific mechanisms or scenarios that meet their requirements. 
For instance, DP-Finder \cite{bichsel2018dp} requires the mechanisms to have a differentiable power on the witness. Therefore, it does not support any mechanism involving arbitrary loops or hash functions, such as RAPPOR \cite{rappor}.
DP-Stochastic-Tester \cite{wilson2020differentially} requires the mechanism's output space to be $\mathbb{R}$, or its confidence interval computation would fail.
$\mathcal{T}_{\text{priv}}$ \cite{dixit2013testing} requires oracle access to the specific output probability, which is hardly possible for mechanisms in reality. StatDP \cite{ding2018detecting} needs to run the curator's mechanism without any noise in one of its auditing steps, and is already confirmed to be unsound and unstable in \cite{bichsel2021dp}. 
In contrast, blackbox auditors do not require any such assumptions and are thereby universally applicable.

Another non-blackbox auditor for DPSGD leverages parametric estimation. Unlike blackbox auditors' non-parametric estimations, this auditor \cite{nasr2023tight} knows the functional form of the curator's type II-type I error tradeoff, and applies parametric estimations on the collected samples to determine its power $\xi$. 
Compared to blackbox auditing, such auditors are limited to mechanisms with definable parametric tradeoff models, e.g. the Gaussian mechanism. They fail when the tradeoff is hard to express in a closed form, e.g. the SVT mechanism.
